\documentclass[12pt]{article}

\usepackage{arxiv}

\usepackage{parskip}
\usepackage[utf8]{inputenc} 
\usepackage[T1]{fontenc}    
\usepackage[hidelinks]{hyperref}
\hypersetup{
	colorlinks=true,       
	linkcolor=blue,         
	citecolor=blue,        
	filecolor=blue,      
	urlcolor=blue           
}
\usepackage{url}            
\usepackage{booktabs}       
\usepackage{amsfonts,amssymb}       
\usepackage{nicefrac}       
\usepackage{microtype}      
\usepackage{xcolor}         
\usepackage{natbib}

\usepackage{algorithm}
\usepackage[noend]{algpseudocode}
\usepackage{amsmath,amstext}
\usepackage{mathrsfs}
\usepackage{bbm}
\usepackage[pdftex]{graphicx} 
\usepackage{multirow}
\usepackage{enumitem} 

\usepackage{amsthm}

\newtheorem{theorem}{Theorem}
\newtheorem{lemma}[theorem]{Lemma}

\newtheorem{proposition}[theorem]{Proposition}
\newtheorem{example}[theorem]{Example}
\newtheorem{definition}[theorem]{Definition}

\usepackage[labelformat=simple]{subcaption}

\usepackage{dsfont}

\newcommand{\seqdata}{\mathcal{SD}} 
\newcommand{\seq}{\mathcal{S}}
\newcommand{\itmset}{\mathbb{I}}
\newcommand{\event}{\ensuremath{e}}
\newcommand{\trfevent}{\bar{\event}}
\newcommand{\eventspace}{\mathbb{E}}
\newcommand{\trfeventspace}{\eventspace}
\newcommand{\numseq}{\ensuremath{N}}
\newcommand{\numevent}{\ensuremath{|\eventspace|}}
\newcommand{\maxseqsize}{\ensuremath{M}} 
\newcommand{\order}{\mathcal{O}}
\newcommand{\prefix}{\mathcal{\check{S}}}
\newcommand{\postfix}{\mathcal{\hat{S}}}
\newcommand{\pat}{\mathcal{P}}
\newcommand{\minthr}{\theta}
\newcommand{\minfreq}{\ensuremath{\minthr \times \numseq}}

\newcommand{\vecmodel}{\mathcal{V}}
\newcommand{\trimodel}{\mathcal{T}}
\newcommand{\btrie}{\mathcal{BT}}
\newcommand{\htrie}{\mathcal{HT}}
\newcommand{\memint}{c^{\text{int}}}
\newcommand{\memvec}{c^{\vecmodel}}
\newcommand{\memtri}{c^{\trimodel}}
\newcommand{\tpath}{{\mathbf{P}}}
\newcommand{\nodeset}{\mathcal{N}}
\newcommand{\arcset}{\mathcal{A}}
\newcommand{\labset}{\mathcal{L}}

\newcommand{\optlen}{j^*}
\newcommand{\layer}{\mathcal{N}}
\newcommand{\rootnode}{{\bf{r}}}
\newcommand{\node}{\ensuremath{n}} 
\newcommand{\freq}{\ensuremath{f}} 
\newcommand{\supp}{\ensuremath{supp}}
\newcommand{\ancest}{\ensuremath{a}}
\newcommand{\transform}{\mathbf{S}}
\newcommand{\transformb}{\transform}
\newcommand{\child}{\ensuremath{chl}}
\newcommand{\sibl}{\ensuremath{sib}}

\begin{document}

\title{Memory-Efficient Sequential Pattern Mining with Hybrid Tries}

\author{Amin Hosseininasab\\
       Warrington College of Business\\
       University of Florida\\
       Gainesville, FL, USA\\
       \texttt{a.hosseininasab@ufl.edu}
              \And
       Willem-Jan van Hoeve \\
       Tepper School of Business\\
       Carnegie Mellon University\\
       Pittsburgh, PA, USA\\
       \texttt{vanhoeve@andrew.cmu.edu}
       \And
       Andre A. Cire\\
       Rotman School of Management \\
       University of Toronto\\
       Toronto, ON, Canada\\
       \texttt{andre.cire@rotman.utoronto.ca} 
       }

\maketitle

\begin{abstract}%
This paper develops a memory-efficient approach for Sequential Pattern Mining (SPM), a fundamental topic in knowledge discovery that faces a well-known memory bottleneck for large data sets. Our methodology involves a novel hybrid trie data structure that exploits recurring patterns to compactly store the data set in memory; and a corresponding mining algorithm designed to effectively extract patterns from this compact representation. Numerical results on small to medium-sized real-life test instances show an average improvement of 85\% in memory consumption and 49\% in computation time compared to the state of the art. For large data sets, our algorithm stands out as the only capable SPM approach within 256GB of system memory, potentially saving 1.7TB in memory consumption.
\end{abstract}


\section{Introduction}

Data volume is growing at an exponential rate \citep{datagrowth}, with modern machine learning data sets often containing trillions of data points \citep{datasetsize}. While supervised machine learning algorithms have thrived by training on such large data sets, unsupervised algorithms face ongoing challenges in scalability due to their memory requirements. In particular, Sequential Pattern Mining (SPM), a prominent topic in unsupervised learning, encounters a well-known memory bottleneck in its two most prevalent algorithms. The Apriori algorithm \citep{agrawal1994fast} suffers from the explosion of candidate patterns that are costly to store in memory, and the prefix-projection algorithm \citep{han2001prefixspan} requires fitting the entire training data set into memory. This has limited extant SPM algorithms to smaller-sized data sets and rendered them impractical for larger ones \citep{pillai2011user}. 

Larger data sets are inherently richer in information, and mining them can uncover intricate patterns that facilitate a deeper understanding of the relationships in data. This includes rare-event patterns and those with long-term dependencies that are more prevalent in larger samples of the population, but may be infrequent in smaller subsets. Such patterns are of interest in numerous applications, such as fraud detection \citep{kim2022sequential}, medical research \citep{ji2012method}, bioinformatics \citep{bechet2012sequential}, and market basket analysis \citep{pillai2011user}, to name a few. Additionally, patterns mined from a larger sample have less variance and are statistically more robust than the same patterns mined from a smaller subset \citep{hamalainen2008efficient}. This highlights the need for memory-efficient SPM algorithms that can handle larger data sets and adapt to their rapidly growing data environment. 

The primary approach to applying SPM algorithms on large data sets is by using more hardware, either on a single computing unit or via a parallelized structure, such as  \citet{gan2019survey,huynh2018efficient,chen2017distributed,yu2019scalable,saleti2019novel,chen2013highly}. However, using additional hardware is costly, requires specialized machinery in the case of parallelization, and is often capped due to technology and other system specifications. Maximizing performance under capped memory is thus a practical use-case that is of interest in SPM \citep{wang2003closet+}, and has been shown to be beneficial in various machine learning techniques such as \citet{pleiss2017memory,gruslys2016memory,si2017memory}. Benefits include, for example, faster data set access and higher time efficiency, reduced overhead and optimized resource utilization, reduced hardware maintenance costs, facilitation of use in dynamic and online data streams, reduced environmental impact, and higher energy efficiency. Most importantly, a memory-efficient algorithm makes SPM available to a broader range of users who are constrained by hardware limitations. 

Motivated by these benefits, this paper aims to enhance the memory efficiency of SPM algorithms, while simultaneously preserving or increasing their time efficiency. We focus on the prefix-projection algorithm, which has been shown to improve over other alternative SPM algorithms in terms of time efficiency \citep{han2001prefixspan}. The memory bottleneck of prefix-projection arises from the necessity to fit the entire data set into memory, prompting a critical examination of how sequential data sets are modeled and stored. The dominant approach is to model the data set using a relational or \emph{vector} model, such as the ones discussed in \citet{fournier2017survey}. An advantage of vector models is their simple structure, leading to a straightforward mining process. Nevertheless, a major disadvantage of vector models is their high memory usage which becomes increasingly inefficient as data sets grow in size. This has led researchers to explore alternative models such as \emph{trie} structures which are known for their concise representation of strings.  

A trie is a graph-based data structure commonly used to store associative arrays or sets of strings in an efficient manner. A major advantage of trie models is their ability to model multiple overlapping subsequences using only a single trie path, which can potentially lead to higher time and memory efficiency in the mining algorithm \citep{mabroukeh2010taxonomy}. For example, trie models of data sets have been shown to provide benefits such as faster item-set mining \citep{han2004mining,pyun2014efficient,borah2018fp}, effective Apriori and candidate pattern storage \citep{ivancsy2005efficient,pyun2014efficient,huang2008general,fumarola2016clofast,antunes2004sequential,wang2006scalable,bodon2003trie,masseglia2000efficient}, effective web access mining \citep{yang2007efficient,pei2000mining,lu2003position}, mining long biological sequences \citep{liao2014dfsp}, up-down SPM \citep{chen2009updown}, incorporating constraints into SPM \citep{masseglia2009efficient,hosseininasab2019constraint,wang2022seq2pat,kadiouglu2023seq2pat}, progressive SPM \citep{huang2006progressive,el2004fs}, and faster SPM \citep{rizvee2020tree}. 

While trie models can provide a more memory-efficient model of the data set, using that model for SPM poses an entirely different challenge. Unlike vector models which have a one-to-one correspondence between their vectors and sequences of the data set, a single path of a trie model may correspond to several sequences. This makes tracking the frequency of patterns nontrivial, requiring additional information to be stored at the nodes of the trie. In fact, the distinguishing factor between the many trie-based approaches in the literature is the information stored at the nodes of their tries and how it is used in the mining algorithm. Unfortunately, all current trie models involve storing a rich set of information at their nodes in favor of time efficiency, often increasing their memory requirements beyond that of vector models. 

For example, in web access mining (a special case of SPM with simplified sequence structures), \citet{pei2000mining} propose to use a hash table of linked nodes to traverse and scan their trie data set model. Using hash tables, the data set is recursively projected onto conditional smaller tries and mined accordingly. The disadvantage of this approach is the memory overhead of the hash table and the memory and time spent constructing the conditional tries. Instead of constructing conditional tries, \citet{lu2003position} propose to store at each node of the trie model an integer position value, and \citet{yang2007efficient} propose to recursively generate sub-header tables in the mining algorithm. Such integer position values grow exponentially with the size of the trie, and similarly, the generation of many sub-header tables leads to higher memory consumption. 

Following works in web access mining, \citet{rizvee2020tree} extend the trie model of a sequential data set to accommodate SPM. This is a challenging task, due to the different structural properties of data sets in SPM compared to web access mining \citep{mabroukeh2010taxonomy}. Their \emph{TreeMiner} algorithm uses a similar idea to hash tables, and stores at each node of the trie a matrix of links that are used to traverse and mine the trie. Although this can improve the time efficiency of the TreeMiner algorithm, the matrix of links grows linearly with the size of the data set and makes the algorithm costly in memory usage.

To improve over the high memory usage of vector and trie models, we begin by introducing a new binary trie model of the data set that stores a constant amount of information at each node. We prove that this information suffices for SPM, and develop a novel algorithm with significantly lower memory requirement than that of \citet{rizvee2020tree}. Furthermore, we prove that our trie model is always asymptotically smaller than a vector model, giving it a major advantage in SPM of large data sets. On the other hand, in data sets where only a few subsequences overlap, such as smaller data sets or ones with longer sequences, the memory overhead of modeling a sequence by a trie path may be higher than its vector representation. 

To improve memory efficiency for such data sets, we build on our binary trie and develop a hybrid trie-vector model of the data set. The idea is to take advantage of the strengths of both trie and vector models and further increase the time and memory efficiency of the subsequent SPM task. In particular, we exploit the fact that data set sequences have high overlap in their initial entries (that is, their prefixes)---which can be effectively modeled via a trie---and low overlap in their latter entries (that is, their suffixes)---which are more efficiently modeled via a vector. Our hybrid model is thus designed to find an effective transition from a trie model of prefixes to a vector model of suffixes that improves memory consumption. 

To mine our hybrid data set structure, we combine our binary trie mining algorithm with a vector-based prefix-projection algorithm to develop a novel hybrid mining algorithm. We experimentally show that our hybrid algorithm outperforms both trie and vector models in time and memory usage. In particular, our hybrid algorithm can model and mine orders of magnitude larger data sets that are 
out of reach for any other SPM algorithm. Although primarily designed to handle large data sets, our algorithm improves time and memory usage when applied to small to medium-sized data sets, showcasing its potential for larger data sets in time efficiency. 

The rest of the paper is organized as follows. We begin by discussing the preliminaries of SPM in \S\ref{sec:prelim}, including vector and trie models of the literature. We then develop our novel binary trie and hybrid models and associated mining algorithms in \S\ref{sec:methodology}. Numerical results on real-life large-size data sets are given in \S\ref{sec:results}, and the paper is concluded in \S\ref{sec:conclusion}.

\section{Preliminaries}
\label{sec:prelim}

This section provides preliminaries on SPM, starting with the definition of vector models of sequential data sets and followed by the definition of trie models. Throughout the paper, and for a thorough space complexity analysis, we consider that a vector of integers uses a constant $\memvec$ memory overhead, and that an integer $j$ has $\order(\log(|j|))$ space complexity \citep{papadimitriou1998combinatorial}. We also discuss the relaxation of the later assumption for \emph{reasonably-sized} integers that can be stored in memory using constant overhead. 

\subsection{Vector Models of the Data Set for SPM}

Let $\eventspace = \left\lbrace \event_1, \dots, \event_{\numevent} \right\rbrace$ be a finite set of literals, representing possible \emph{events} or items within an application of interest. An \emph{itemset} $\itmset = \left\lbrace \event_1,\dots,\event_{\left|\itmset\right|} \right\rbrace$ is a set of events such that $\itmset \subseteq \eventspace$. Although events of an itemset can be of any order, they are generally assumed to satisfy a monotone property and ordered accordingly \citep{han2001prefixspan}. A \emph{sequence} $\seq_i = \left\langle \itmset_i^1, \dots, \itmset_i^{L_i} \right\rangle$ is an ordered list of $L_i$ itemsets, with size $|\seq_i| = \sum_{j=1}^{L_i} |\itmset_i^j|$. An event $\event \in \eventspace$ can occur at most once in an itemset, but the same itemset can occur multiple times in a sequence. Sequences may be equivalently represented in the event space $\seq_i=\left\langle  \trfevent_1 , \dots, \event_j, \trfevent_{j+1},  \dots, \event_{|\seq_i|} \right\rangle$, with the first event of any itemset indicated by an accented event $\trfevent$. This notation is adopted throughout this paper.

A \emph{sequential data set} $\seqdata$ is a list of $\numseq$ sequences $\seqdata = \left\langle \seq_1, \dots, \seq_\numseq \right\rangle$, with the size of its largest sequence denoted by $\maxseqsize = \max_{i \in \lbrace1,\dots,\numseq\rbrace} |\seq_i|$. Example~\ref{ex:seqdataset} describes a small instance of a sequential data set given in Table~\ref{tab:vector}, which we use as a running example.  
\begin{example}\label{ex:seqdataset}
    Table~\ref{tab:vector} gives a vector model of a sequential data set that includes $\numseq = 20$ sequences $\seq_1, \dots, \seq_{20}$ with events $\eventspace = \{a,b,c\}$, and maximum sequence length $\maxseqsize = 6$. Sequences are given in itemset form, but can be equivalently represented in their event space. For instance, sequence $\seq_1$ is a list of (ordered) itemsets $\langle \lbrace a, b \rbrace, \lbrace a \rbrace, \lbrace b \rbrace \rangle$ that can be equivalently represented as $\langle \overline{a}, b, \overline{a},\overline{b} \rangle$.
\end{example}

\begin{table}
\centering
    \caption{Vector model of a sequential data set. }  
\label{tab:vector}
\begin{tabular}{cl|cl}
   \toprule
    $\seq_1$ & $\langle \lbrace a, b \rbrace, \lbrace a \rbrace, \lbrace b \rbrace \rangle$ &
    $\seq_{11}$ & $\langle \lbrace a, c \rbrace, \lbrace a, c \rbrace \rangle$ \\ 
    $\seq_2$ & $\langle \lbrace a, b, c \rbrace, \lbrace c \rbrace \rangle$ & 
    $\seq_{12}$ & $\langle \lbrace a \rbrace \rangle$ \\
    $\seq_3$ & $\langle \lbrace a \rbrace, \lbrace b \rbrace \rangle$ & 
    $\seq_{13}$ & $\langle \lbrace a, b \rbrace, \lbrace a \rbrace, \lbrace b, c \rbrace \rangle$ \\
    $\seq_{4}$ & $\langle \lbrace a, c \rbrace, \lbrace a \rbrace \rangle$ & 
    $\seq_{14}$ & $\langle \lbrace a, c \rbrace, \lbrace a, c \rbrace, \lbrace a, c \rbrace \rangle$ \\
    $\seq_{5}$ & $\langle \lbrace a, b, c \rbrace \rangle$ & 
    $\seq_{15}$ & $\langle \lbrace a, c \rbrace, \lbrace a \rbrace \rangle$\\
    $\seq_6$ & $\langle \lbrace a, b \rbrace \rangle$ & 
    $\seq_{16}$ & $\langle \lbrace a, b \rbrace, \lbrace a \rbrace, \lbrace b \rbrace \rangle$ \\
    $\seq_7$ & $\langle \lbrace a, b, c \rbrace \rangle$ & 
    $\seq_{17}$ & $\langle \lbrace a, b \rbrace, \lbrace a \rbrace, \lbrace b \rbrace, \lbrace a, c \rbrace \rangle$\\
    $\seq_{8}$ & $\langle \lbrace a, b, c \rbrace, \lbrace c \rbrace \rangle$ & 
    $\seq_{18}$ & $\langle \lbrace a, c \rbrace, \lbrace a , c \rbrace, \lbrace b, c \rbrace \rangle$\\
    $\seq_{9}$ & $\langle \lbrace a, c \rbrace, \lbrace a \rbrace \rangle$ & 
    $\seq_{19}$ & $\langle \lbrace a, b, c \rbrace, \lbrace c \rbrace \rangle$\\
    $\seq_{10}$ & $\langle \lbrace a, b \rbrace \rangle$ & 
    $\seq_{20}$ & $\langle \lbrace a, b \rbrace, \lbrace a \rbrace \rangle$\\
\bottomrule  
\end{tabular}
\end{table}

A sequence $\seq_{i'}$ is said to be a \emph{subsequence} of another sequence $\seq_i$, denote $\seq_{i'} \sqsubseteq \seq_i$, if there exist integers $k_1 < \dots < k_{L_{i'}}$ such that $\itmset_{i'}^j \subseteq \itmset_i^{k_j}$ for all $j = 1,\dots,L_{i'}$. A \emph{prefix} $\prefix_i$ is a contiguous subsequence of $\seq_i$, with $k_1 = 1$ and $k_{j + 1} = k_j + 1$ for all $j = 1, \dots, |\prefix_i| - 1$. Similarly, a \emph{postfix} $\postfix_i$ is a contiguous subsequence of $\seq_i$, with $k_1 = |\seq_i| - |\postfix_i| + 1$ and $k_{j + 1} = k_j + 1$ for all $j = 1, \dots, |\postfix_i| - 1$. 

The SPM task involves finding the set of \emph{frequent patterns} within a data set $\seqdata$. A pattern $\pat$ is a subsequence that satisfies $\pat \sqsubseteq \seq_i$ for at least one sequence $\seq_i \in \seqdata$. The \emph{support} $\supp(\pat) \in \mathbb{Z}_{> 0}$ of a pattern $\pat$ is the number of distinct sequences in $\seqdata$ for which $\pat \sqsubseteq \seq_i$. A pattern is considered frequent if $\supp(\pat) \ge \minfreq$, where $0<\minthr\le 1$ is a user-defined \emph{minimum support threshold}. 

Frequent patterns are generally found iteratively, where at each iteration a frequent pattern $\pat$ is extended by a single event $\event \in \trfeventspace$, to $\langle \pat, \event \rangle$, and checked to satisfy $\supp(\langle\pat, \event \rangle) \ge \minfreq$. Pattern extensions are classified into an \emph{itemset extension} or a \emph{sequence extension}. In an itemset extension, pattern $\pat$ is extended to $\langle \pat, \event \rangle$, extending its last itemset $\itmset^{|\pat|}$ by a single event. In a sequence extension, a pattern is extended to $\langle \pat, \trfevent \rangle$, extending $\pat$ by a new itemset $\itmset = \lbrace \trfevent \rbrace$. Example~\ref{ex:pattern} demonstrates a frequent pattern in our running example. 
\begin{example}
	\label{ex:pattern}
    Given a support threshold of $\minthr = 0.2$, an example frequent pattern in Table~\ref{tab:vector} is $\pat = \langle \lbrace a, b \rbrace \rangle$. The pattern has support $\supp(\pat) = 12 \ge 4$, as it is a subsequence of sequences $\seq_1, \seq_2, \seq_{5}, \seq_6 - \seq_8, \seq_{10}, \seq_{13}, \seq_{16}, \seq_{17}, \seq_{19}, \seq_{20}$ with $k_1 =1, k_2 = 2$ for all sequences. An example sequence extension of $\pat$ is $\pat' = \langle \lbrace a, b \rbrace, \lbrace c \rbrace \rangle$ with support $\supp(\pat') = 5$ (sequences $\seq_2, \seq_8, \seq_{13}, \seq_{17}, \seq_{19}$). An example itemset extension is $\pat'' = \langle \lbrace a, b, c \rbrace \rangle$ with support $\supp(\pat'') = 5$ (sequences $\seq_2, \seq_5, \seq_{7}, \seq_{8}, \seq_{19}$).
\end{example}

The literature generally models sequential data sets via vectors, where each sequence $\seq_i \in \seqdata$ is stored in memory using a single vector. Current state-of-the-art SPM algorithms that use vector models store their entire representation of the data set in memory, which can be costly in terms of memory usage. In particular, Lemma~\ref{lem:vecspace} gives the worst-case space complexity of a vector-based SPM algorithm.
\begin{lemma} \label{lem:vecspace}
    The worst-case space complexity of a vector-based SPM algorithm is $\order\left(\numseq\maxseqsize\log(\numseq) \right)$, and $\order\left(\numseq\maxseqsize \right)$ for reasonably-sized integers $\numseq$.
\end{lemma}
\begin{proof}
    A sequential data set $\seqdata$ uses one vector per sequence $\seq_i \in \seqdata$ and stores an integer $j \le \numevent$ per event $\event \in \seq_i$. Assuming that $\numevent \le \numseq$ and $\maxseqsize \le \numseq$ (which generally hold in practice), the worst-case space complexity of a vector model is $\order\left(\numseq\memvec + \numseq\maxseqsize\log(\numevent) \right) = \order\left(\numseq\maxseqsize\log(\numevent) \right)$. 

    The most memory-efficient and basic SPM algorithm involves pseudo-projection \citep{han2001prefixspan}. Pseudo-projection stores a sequence ID-integer position pair $(i,j)$ (with $i \le \numseq$ and $j \le \maxseqsize$) for at most all $\seq_i \in \seqdata$ and positions $\maxseqsize$. The worst-case space complexity of a vector-based SPM algorithm is thus $\order(\numseq\maxseqsize\log(\maxseqsize) + \numseq\maxseqsize\log(\numseq)) = \order(\numseq\maxseqsize\log(\numseq))$. 

    Putting the two together, the worst-case space complexity of vector-based SPM algorithms is $\order(\numseq\maxseqsize\log(\numevent) + \numseq\maxseqsize\log(\numseq) ) = \order\left(\numseq\maxseqsize\log(\numseq)\right)$. Reasonably sized integers $\numseq$ consume constant memory, and reduce the complexity to $\order\left(\numseq\maxseqsize\right)$
\end{proof}
As shown in Lemma~\ref{lem:vecspace}, the memory consumption of vector models and algorithms grows linearly with $\numseq$---for reasonably sized values of $\numseq$. In practice, this can be highly memory-consuming, for example, for the large-size data sets of \citet{datasetsize,criteo,10002015global} that include billions to trillions of sequences. For such large data sets, vector-based SPM algorithms cannot fit their data set representation into system memory, which consequently prevents them from performing their mining task. To improve on this memory deficiency, we next examine trie models of the data set, which have the potential to conserve memory by representing multiple sequences of the sequential data set using a single path.

\subsection{Trie Models of the Data Set for SPM}
\label{sec:gentriemodels}

For a sequential data set $\seqdata$, let $\trimodel := (\nodeset, \arcset, \labset)$ be its \emph{labeled} trie model with node set $\nodeset$, arc set $\arcset$, and a set of labels $\labset$. The node set $\nodeset$ can be partitioned into $\maxseqsize+1$ subsets $\layer_0, \dots, \layer_{\maxseqsize}$, referred to as layers. Layer $\layer_0 := \{\rootnode\}$ is a singleton containing auxiliary root node $\rootnode$, and the remaining $j=1,\dots,\maxseqsize$ layers model the events of sequences $\seq_i \in \seqdata$. Accordingly, each node $\node \in \nodeset \setminus \{\rootnode\}$ is associated with an \emph{event label} $\event_\node \in \labset$, which denotes the event literal $\event \in \trfeventspace$ modeled by node $\node$; and an \emph{itemset label} $\itmset_{\node}$, which denotes the position $j$ of itemset $\itmset_i^j$ in the sequence $\seq_i$ \emph{modeled} by path $\tpath$. 

Using event and itemset labels $\event_{\node}$, a trie path $\tpath = (\rootnode, \node_1, \dots, \node_{|\tpath|})$ models the sequence $\seq_i = \langle \event_{\node_{1}}, \dots, \event_{\node_{|\tpath|}}\rangle$ belonging to itemsets $\langle \itmset_{\node_{1}}, \dots, \itmset_{\node_{|\tpath|}}\rangle$. For notation convenience, we specify the sequence $\seq_i$ modeled by a trie path $\tpath$ using a \emph{transformation function} $\transform(\tpath) = \seq_i$. 
 
The main advantage of trie models is their ability to model all overlapping prefixes $\prefix_i$ of sequences in $\seqdata$ using only a single path $\tpath: \transform(\tpath) = \prefix_i$. Accordingly, nodes $\node \in \nodeset \setminus \{\rootnode\}$ are associated with a positive integer \emph{frequency label} $\freq_\node \in \labset$ that denotes the number of prefixes $\prefix_i$ of sequences $\seq_i \in \seqdata$ modeled by path $\tpath = (\rootnode, \dots, \node)$. Example~\ref{ex:seqtopath} illustrates the general trie model of the data set in our running example.

\begin{figure}
	\centering
	\includegraphics[scale=0.6]{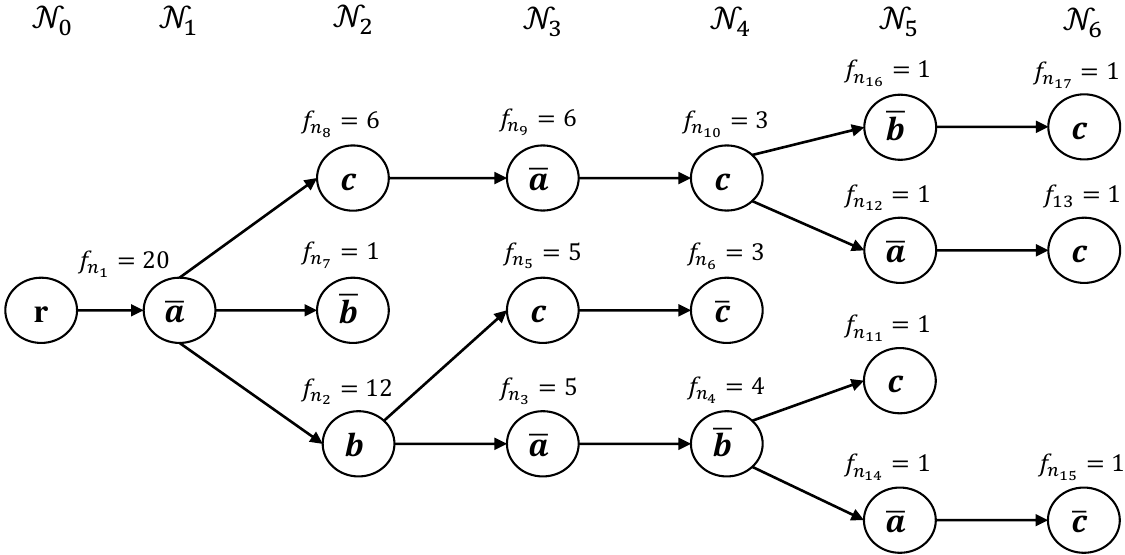}
     \captionsetup{justification=centering}
	\caption{General trie model $\trimodel$ of the data set in Table~\ref{tab:vector}. Although the general trie structure correctly models all sequences of the data set and their frequencies, its labels are insufficient for SPM.}
	\label{fig:gentrie}
\end{figure} 

\begin{example}\label{ex:seqtopath}
    Figure~\ref{fig:gentrie} depicts the trie model of the data set given in Table~\ref{tab:vector}. Event labels are displayed inside each node, frequency labels are given above each node, and node layers are given above each column of nodes. The trie models the 20 sequences of the sequential data using 6 maximal paths and a total of $|\nodeset| = 17$ nodes. Note that a vector model uses 20 vectors and a total of $\sum_{\seq_i \in \seqdata} |\seq_i| = 72$ units of memory to store the same data set. This is considerably less memory efficient even for our small running example. Although the trie correctly models all sequences of the data set and their frequencies, its labels are insufficient for SPM.  
\end{example}

Forgoing the differences between web access mining and SPM, the trie models of \citet{yang2007efficient,pei2000mining,lu2003position,rizvee2020tree} are identical in their node and arc set $(\nodeset, \arcset)$. In particular, the contribution of the trie models in the literature is the additional labels in $\labset$ that are required to perform web access mining or SPM (see, for example, \citet{pei2000mining,lu2003position,yang2007efficient,rizvee2020tree}). These label sets are critical in the mining process, and in the case of web access mining, do not generalize to SPM. This is due to the inherent difference between web access mining and SPM, where web access algorithms cannot distinguish between itemset and sequence extensions \citep{mabroukeh2010taxonomy}.

\begin{figure}[t!]
	\centering
	\includegraphics[scale=0.6]{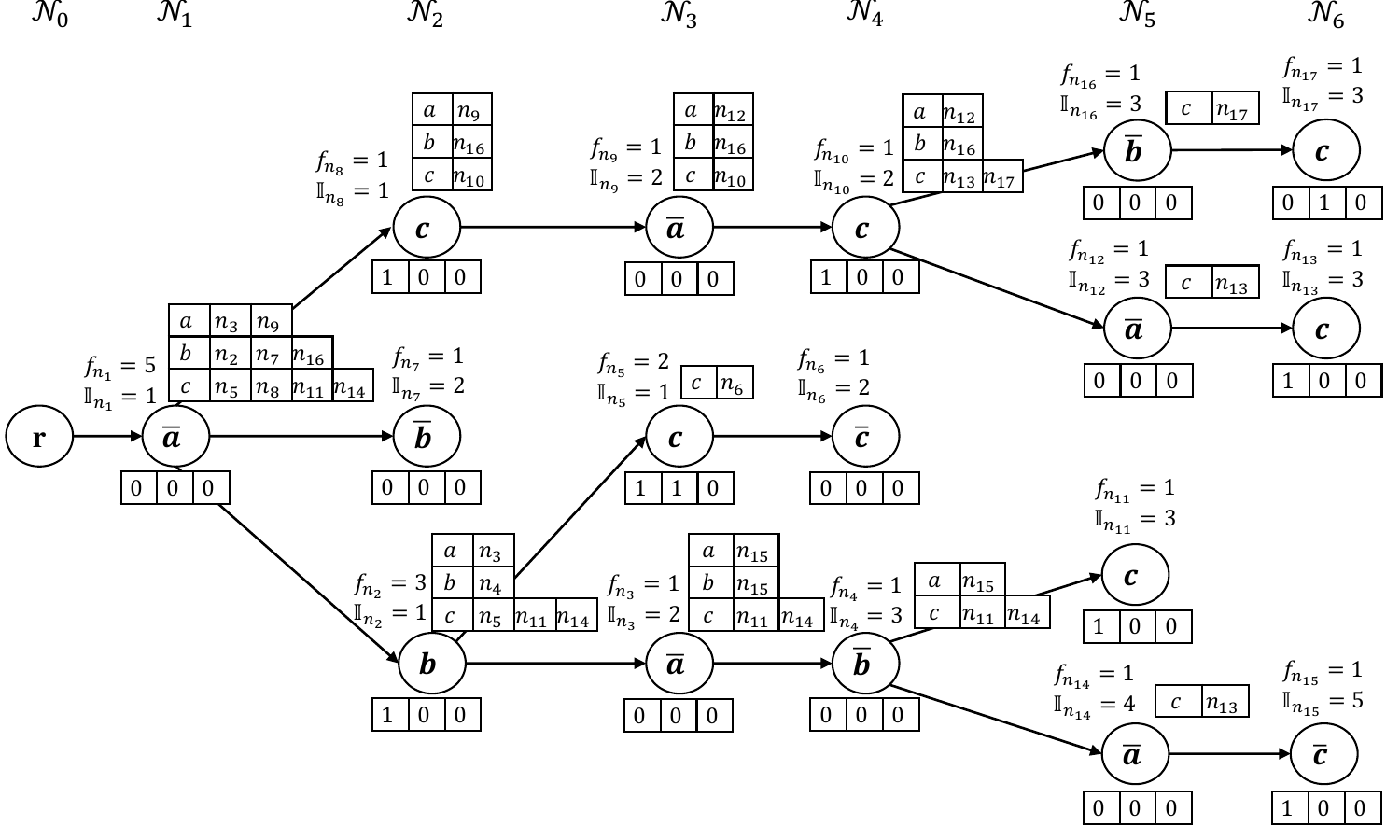}
     \captionsetup{justification=centering}
	\caption{Trie model of TreeMiner for the data set of Table~\ref{tab:vector}. The trie is updated with itemset labels $\itmset$, next-links displayed as a matrix adjacent to each node, and the parent-info bitset displayed as a vector under each node. TreeMiner mines the trie by traversing it using next-links and determining valid extensions using parent-info and itemset labels.}
	\label{fig:treeminer}
\end{figure} 

In order to model and mine data sets in SPM, \citet{rizvee2020tree} expand the label set $\labset$ of each node $\node \in \nodeset$ by two additional labels. The first ``next-links'' label is a matrix of node pointers, where each row corresponds to an event $\event \in \eventspace$, and each column points to the first node $\node': \event_{\node'} = \event$ of each path $(\node, \dots, \node')$ spanning from node $\node$. The second ``parent-info'' label is a bitset that determines which events $\event \in \eventspace$ fall into the same itemset as event $\event_\node$ on the path $(\rootnode, \dots, \node)$. These labels are used in a tailored mining algorithm, TreeMiner, to mine all patterns. Figure~\ref{fig:treeminer} shows the trie model of TreeMiner for our running example.

TreeMiner traverses its trie model using the next-link matrices and mines patterns using the information stored in the parent-info and itemset labels. Although next-links can enable faster traversal of the trie and SPM is possible using the parent-info and itemset labels, storing them at each node of the trie model is costly in memory usage. In particular, Lemma~\ref{lem:spacetminer} gives the worst-case space complexity of the TreeMiner algorithm. 
\begin{lemma} \label{lem:spacetminer}
    The worst-case space complexity of TreeMiner is $\order\left(\min\lbrace \numseq, \numevent^{\maxseqsize}\rbrace\maxseqsize\numevent^2\log(\numseq) \right)$, and \newline $\order\left(\min\lbrace \numseq, \numevent^{\maxseqsize}\rbrace\maxseqsize\numevent^2 \right)$ for a reasonably-sized $\numseq$.
\end{lemma}
\begin{proof}
    A node $\node \in \trimodel$ of the trie model of TreeMiner stores integer values $\freq_\node$, $\event_\node$, $\itmset_\node$, which are all bounded by $\order(\log(\numseq))$, a next-link matrix bounded by $\order((\numevent + 1)\memvec + \numevent^2 \log(\numseq)))$, a parent-info bitset bounded by $\order(\memvec + \numevent)$, and a children vector bounded by $\order\left(\memvec + \numevent\log(\numseq)\right)$. The space complexity of a TreeMiner node is thus $\order\left(\numevent^2\log(\numseq)\right)$.

    By definition, a node $\node$ of a trie model has at most $2\numevent$ children, an event $\event \in \eventspace$ representing an itemset extension and an event $\trfevent \in \eventspace$ representing a sequence extension. The maximum number of children for a trie layer $\layer_j$ is thus $\left(2\numevent\right)^{|\layer_j|}$. As $|\layer_0|=1$, the maximum number of nodes for any layer is $\left(2\numevent\right)^\maxseqsize$. On the other hand, we can have at most $\numseq$ nodes at any layer of a trie model of a sequential data set $\seqdata$, where each sequence $\seq_i \in \seqdata$ is modeled by exactly one path. The maximum number of nodes in a trie model is thus $\order\left(\min\lbrace \numseq, \numevent^{\maxseqsize}\rbrace\maxseqsize\right)$. For TreeMiner, this gives the overall worst-case memory complexity of its trie model as $\order\left(\min\lbrace \numseq, \numevent^{\maxseqsize}\rbrace\maxseqsize\numevent^2\log(\numseq)\right)$.

    For its mining algorithm, TreeMiner stores one positional integer $j \le \maxseqsize$ for at most every node in its trie, bounding its worst-case space complexity by $\order\left(\min\lbrace \numseq, \numevent^{\maxseqsize}\rbrace\maxseqsize\log(\numseq)\right)$. Putting the two together, gives the overall space complexity of TreeMiner as $\order\left(\min\lbrace \numseq, \numevent^{\maxseqsize}\rbrace\maxseqsize\numevent^2\log(\numseq) \right)$. Reasonably sized integers $\numseq$ consume constant memory and reduce the complexity to \newline $\order\left(\min\lbrace \numseq, \numevent^{\maxseqsize}\rbrace\maxseqsize\numevent^2 \right)$.
\end{proof}
The memory efficiency of TreeMiner is highly dependent on the number of events $\numevent$ and maximum sequence length $\maxseqsize$. Consequently, TreeMiner may use more memory than a vector model when either value is high, and the data set does not contain many overlapping sequences. We indeed observe this in a number of data sets in our numerical results. For a more memory-efficient SPM algorithm, we propose two novel approaches in the following section. 

\section{Binary and Hybrid Tries for Memory Efficient SPM}\label{sec:methodology}

We introduce our binary trie $\btrie$ in \S\ref{sec:btrie}, and its corresponding mining algorithm in \S\ref{sec:btminer}. We build on $\btrie$ to develop a hybrid trie $\htrie$ and its corresponding mining algorithm in \S\ref{sec:htrie}. 

\subsection{Binary Tries} \label{sec:btrie}

A binary trie $\btrie$ is a doubly chained implementation of a trie $\trimodel$ \citep{sussenguth1963use}, with a novel addition to its set of labels $\labset$. The binary structure of the trie is intended to reduce memory consumption in practice by avoiding the use of vectors to store a node's children. Instead, a node $\node \in \btrie$ is associated with at most one child node $\child(\node)$ and one sibling node $\sibl(\node)$. A child node $\child(\node)$ models the first child of node $\node$, with the remaining children modeled as contiguous siblings of $\child(\node)$. 

The main contribution of $\btrie$ is its updated set of labels which consumes a constant amount of memory. In addition to event, frequency, and itemset labels of a general trie model, $\btrie$ stores at each node $\node \in \nodeset$ an additional \emph{ancestral label} $\ancest_\node \in \labset$, given by Definition~\ref{def:ancestor}.
\begin{definition} \label{def:ancestor}
    The ancestor label $\ancest_\node$ of a node $\node \in \btrie$ is defined as:
    \begin{equation*}
        \ancest_\node = 
        \left\{ 
        \begin{array}{ll}
        \multirow{2}{*}{$\node'$} & \textrm{if } \exists \node': (\node',\dots,\node) \in \btrie, \event_{\node'} = \event_{\node}, \textrm{ and } \\ &\event_{\node''} \neq \event_{\node} \forall \node'' \in (\node',\dots,\node): \node'' \neq \node, \node'' \neq \node',\\
        \rootnode & \textrm{otherwise.}
        \end{array}\right.
    \end{equation*}
\end{definition}
Intuitively, the ancestor label of a node $\node$ tracks the first occurrence of event $\event_\node$ prior to node $\node$, on the path $(\rootnode, \dots, \node) \in \btrie$. As we later show, ancestor labels are sufficient to effectively mine all patterns from a trie model, and can be efficiently generated during its construction. Example~\ref{ex:BDTrie} illustrates the binary-trie model for our running example.
\begin{figure}
    \centering
    \includegraphics[width=0.7\textwidth]{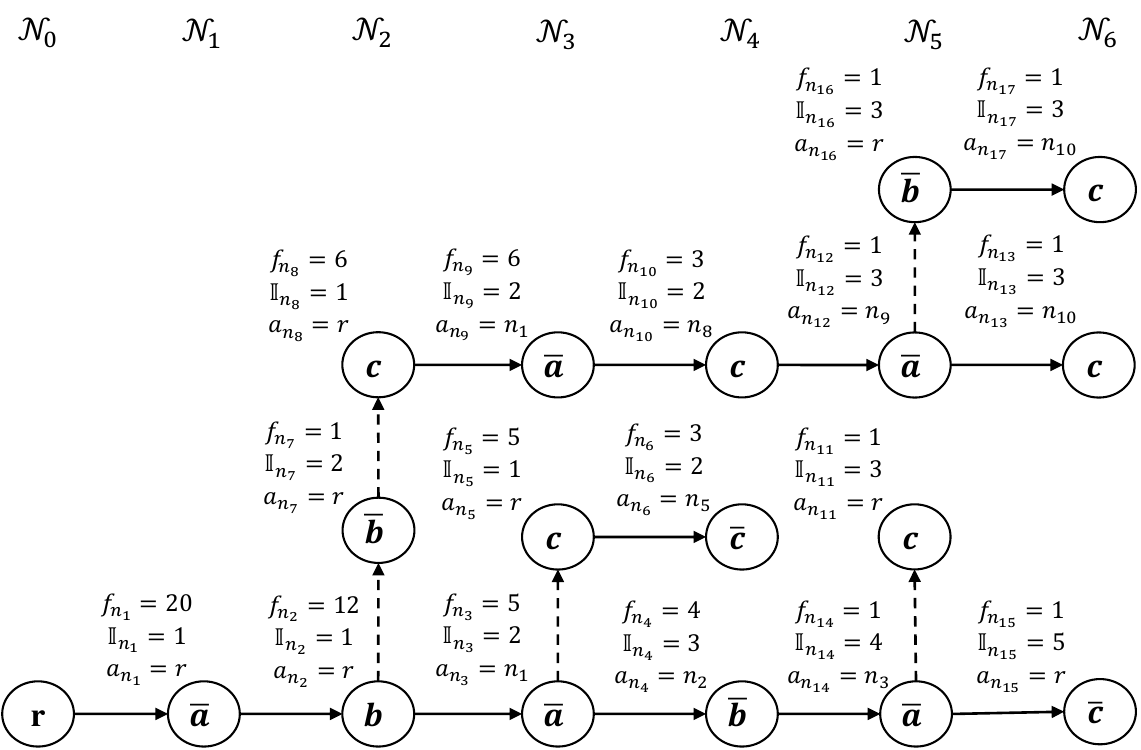}
        \captionsetup{justification=centering}
    \caption{The $\btrie$ model of the data set in Table~\ref{tab:vector}. Sibling nodes are connected by dashed arcs. The trie is updated with itemset labels $\itmset_\node$ and ancestor labels $\ancest_\node$, which are sufficient for SPM.}
    \label{fig:BDTrie}
\end{figure}
\begin{example}\label{ex:BDTrie}
    Consider the binary trie model of our running example given in Figure~\ref{fig:BDTrie}. The nodes $\node_2, \node_7, \node_8$ are all siblings and model the children of their parent node $\node_1$. Similarly, nodes $\node_3$ and $\node_5$ are siblings with parent node $\node_2$. The ancestor of node $\node_7$ is $\ancest_{\node_7} = \rootnode$ as there is no node $\node': \event_{\node'} = \event_{\node_7}$ on the path $(\rootnode, \node_1, \node_7)$. The ancestor of node $\node_{14}$ is $\ancest_{\node_{14}} = \node_3$, as node $\node_3$ is the first prior node to $\node_{14}$ on the path $(\rootnode, \node_1, \node_2, \node_3, \node_4, \node_{14})$ with the same item $\event_{\node_{14}} = \event_{\node_3}$.
\end{example}


Construction of $\btrie$ involves two scans of $\seqdata$. In the first scan, we follow the works of \citet{pei2000mining} and \citet{lu2003position} and perform support-based filtering on the data set. Support-based filtering involves removing any infrequent event $\event \in \eventspace: \supp(\event) < \minfreq$ from $\seqdata$. Such infrequent events cannot be part of any frequent pattern due to the antimonotone property of $\supp(\pat)$, and thus can be removed without affecting the generation of frequent patterns. In the next scan, $\btrie$ is constructed by iterating over the sequences $\seq_i \in \seqdata$. In each iteration, a sequence $\seq_i$ is modeled by increasing the frequency values $\freq_\node$ for any node $\node \in \tpath: \transformb(\tpath) = \seq_i$ constructed in previous iterations, or by creating a new node if no such node exists in $\btrie$. The complete procedure is given in Algorithm~\ref{alg:BDTries}.
\begin{algorithm}[t!]
	\caption{Construction of $\btrie$.} \label{alg:BDTries}
	\begin{algorithmic}[1]
        \State Filter the data set and remove all events $\event \in \eventspace: \supp(\event) < \minfreq$.
        \State Initialize $\btrie$ with $\nodeset := \{ \rootnode \}$, $\arcset := \emptyset$
        \State Let $\node = \rootnode$
        \For{each $\seq_i \in \seqdata$}
            \State Let Anct$(\event) = \rootnode$ for all $\event \in \eventspace$.
            \State Let ItmSet = 0.
            \For{each $e_j \in \seq_i$}
                \If{node $\node$ has a child $\node' = \child(\node)$}
                    \While{$\event_{\node'} \neq \event_j$ and $\node'$ has a sibling node $\node'' = \sibl(\node')$}
                    \State Let $\node = \node'$,
                    \State Let $\node' = \node''$.
                    \EndWhile
                    \If{$\event_{\node'} = \event_j$}
                        \State Update $\freq_{\node'} = \freq_{\node'} + 1$, and set $\node = \node'$.
                    \Else
                        \State Add $\node': \event_{\node'} = \event_j, \freq_{\node'} = 1, \ancest_{\node'} = \text{Anct}(\event_j), \itmset_{\node'} = \text{ItmSet}$ to $\nodeset_j$.
                        \State Add sibling arc $(\node,\node')$ to $\arcset$.
                    \EndIf
                \Else 
                    \State Add $\node': \event_{\node'} = \event_j, \freq_{\node'} = 1, \ancest_{\node'} = \text{Anct}(\event_j), \itmset_{\node'} = \text{ItmSet}$ to $\nodeset_j$.
                    \State Add child arc $(\node,\node')$ to $\arcset$.
                \EndIf
                \If{$\event_j = \trfevent_j$}
                    \State Update ItmSet = ItmSet + 1.
                \EndIf
                \State Update Anct$(\event_j) = \node'$.
            \EndFor
        \EndFor
		\State \textbf{return} $\btrie$. 
	\end{algorithmic}
\end{algorithm}

\subsection{The $\btrie$Miner Algorithm}\label{sec:btminer}

Given a data set $\seqdata$, the SPM task involves finding all patterns $\pat$ such that $\supp(\pat) \ge \minfreq$. In prefix-projection, the algorithm is initialized by frequent patterns containing a single event, that is, $\pat = \langle \trfevent \rangle: \event \in \trfeventspace, \supp(\event) \ge \minfreq$. At each subsequent iteration, a pattern $\pat$ is extended by a single event $\event \in \trfeventspace$ to $\langle \pat, \event \rangle$, and checked to satisfy $\supp(\langle\pat, \event \rangle) \ge \minfreq$. 

To mine a $\btrie$ model of the data set, we first define $\nodeset(\pat)$ as the set of terminal nodes $\lbrace\node\rbrace$ of all minimal paths $\tpath = (\rootnode, \dots, \node) \in \btrie: \pat \sqsubseteq \transformb(\tpath)$. Paths $\tpath$ are minimal in that $\forall \node' \in (\rootnode, \dots, \node): \node' \neq \node$, we have $\pat \not\sqsubseteq \transformb\left((\rootnode, \dots, \node')\right)$. Proposition~\ref{prop:minprefix} proves that node sets $\nodeset(\event)$ model all minimum-sized prefixes $\prefix_i \sqsubseteq \seq_i: \pat \sqsubseteq \prefix_i$ for all $\seq_i \in \seqdata$.
\begin{proposition} \label{prop:minprefix}
     Node set $\nodeset(\pat)$, as defined above, models all minimum-sized prefixes $\prefix_i \sqsubseteq \seq_i: \pat \sqsubseteq \prefix_i$ for all $\seq_i \in \seqdata$. 
\end{proposition}
\begin{proof}
    The proof follows from the minimal property of paths $\tpath = (\rootnode, \dots, \node) : \node \in \nodeset(\pat)$. Assume by contradiction that the prefix $\prefix_i: \pat \sqsubseteq \prefix_i$ modeled by a minimal path $\tpath: \transformb(\tpath) = \prefix_i$ is not of minimum size. Then there must exist another prefix $\prefix'_i: \prefix'_i \sqsubset \prefix_i, \pat \sqsubseteq \prefix'_i$. Let $\tpath' = (\rootnode, \dots, \node')$ be the trie path that models prefix $\prefix'_i$. As trie paths are unique and $\prefix'_i \sqsubset \prefix_i$, we have $\tpath = \langle \tpath', \dots, \node\rangle$, a contradiction to the minimality of path $\tpath$. As node set $\nodeset(\pat)$ contains all nodes $\node: \pat \sqsubseteq \transformb((\rootnode, \dots, \node))$, then it models all prefixes $\prefix_i \sqsubseteq \seq_i: \pat \sqsubseteq \prefix_i$ for all $\seq_i \in \seqdata$.
\end{proof}

Our mining algorithm, $\btrie$Miner, initializes with the set of pattern-node-set pairs $(\pat = \lbrace \event \rbrace, \nodeset(\pat))$ for all events $\event \in \eventspace$. This can be done effectively by tracking the first occurrence of events $\event \in \eventspace$ of a sequence $\seq_i \in \seqdata$ during the construction of $\btrie$. In the next steps, $\btrie$Miner iteratively takes a pattern-node-set pair $(\pat = \lbrace \event \rbrace, \nodeset(\pat))$, and attempts to construct sets $\nodeset(\langle \pat, \event \rangle)$ by following Proposition~\ref{prop:mine} 
\begin{proposition}\label{prop:mine}
    Given a node set $\nodeset(\pat)$, node set $\nodeset(\langle \pat, \event \rangle)$ is constructed by finding all minimal paths $(\node, \dots, \node') \in \btrie: \event_{\node'} = \event$, for all nodes $\node \in \nodeset(\pat)$. 
\end{proposition}
\begin{proof}
    By Proposition~\ref{prop:minprefix}, all paths $(\rootnode, \dots, \node): \node \in \nodeset(\pat)$ model all minimum-sized prefixes $\prefix_i\sqsubseteq \seq_i: \pat \sqsubseteq \prefix_i$ for all $\seq_i \in \seqdata$. As paths $(\node, \dots, \node')$ are minimal by definition and have $\event_{\node'} = \event$, then paths $(\rootnode, \dots, \node')$ are also minimal and satisfy $\langle \pat, \event \rangle \sqsubseteq \transformb\left((\rootnode, \dots, \node')\right)$. Finding all such paths for all nodes $\node \in \nodeset(\pat)$ thus constructs $\nodeset(\langle \pat, \event \rangle)$.
\end{proof}
Finding minimal paths $(\node, \dots, \node')$ in a trie is a chalenging task, with extant trie-based SPM algorithms requiring a rich set of information to be stored at their nodes. For example, TreeMiner requires traversing the tree using next-links and using parent info information to determine valid pattern extensions. For $\btrie$Miner, the process involves a depth-first-search of the sub-trie rooted at node $\node$ and finding nodes $\node'$ according to Theorem~\ref{thm:ancestor}. 
\begin{theorem}\label{thm:ancestor}
    Let $\node'$ be a node traversed on a path $\tpath = (\node, \dots, \node')$ rooted at a node $\node \in \nodeset(\pat)$. Path $(\node, \dots, \node')$ is minimal for the construction of set $\nodeset(\langle \pat, \event_{\node} \rangle)$ if and only if:
    \begin{itemize}
        \item For a sequence extension $\langle \pat, \trfevent_{\node'} \rangle$ we have $\itmset_{\node'} \neq \itmset_{\node}$ and $\itmset_{\ancest_{\node'}} \le \itmset_{\node}$,
        \item For an itemset extension $\langle \pat, \event_{\node'} \rangle$ we have 
        \begin{itemize}
            \item $\itmset_{\node'} = \itmset_{\node}$, or
            \item $\forall \event_j \in \itmset^{|\pat|}, \exists \node'' \in \tpath: \event_{\node''} = \event_j, \itmset_{\node''} = \itmset_{\node'}$ and 
            \begin{itemize}
                \item $\itmset_{\ancest_{\node'}} < \itmset_{\node}$, or
                \item $\forall \node''' \in \tpath: \event_{\node'''} = \event_{\node'}$ we have $\node''' \notin \nodeset\left(\langle \pat, \event_{\node'} \rangle\right)$.
            \end{itemize}
        \end{itemize}
    \end{itemize}
\end{theorem}
\begin{proof}
    For a sequence extension $\langle \pat, \trfevent_{\node'} \rangle$, assume by contradiction that $\tpath = (\node, \dots, \node')$ is minimal, but $\itmset_{\node'} = \itmset_{\node}$ or $\itmset_{\ancest_{\node'}} > \itmset_{\node}$. If $\itmset_{\node'} = \itmset_{\node}$ then node $\node'$ belongs to the same itemset as node $\node$ and cannot be used for a sequence extension, a contradiction. If $\itmset_{\ancest_{\node'}} > \itmset_{\node}$, then we must have $\tpath = (\node, \dots, \ancest_{\node'}, \dots, \node')$, and consequently $\langle \pat, \trfevent_{\node'} \rangle \sqsubseteq \transformb\left((\node, \dots, \ancest_{\node'})\right)$, contradicting the minimality of $\tpath$. 

    For the converse, assume by contradiction that $\itmset_{\node'} \neq \itmset_{\node}$ and $\itmset_{\ancest_{\node'}} \le \itmset_{\node}$, but that path $\tpath$ is not minimal. Then path $\tpath$ must be of the form $\tpath = (\node, \dots, \node'', \dots, \node')$, such that $\tpath' = (\node, \dots, \node''): \langle \pat, \trfevent_{\node'} \rangle \sqsubseteq \transformb\left((\node, \dots, \ancest_{\node''})\right)$ is minimal. As $\node''$ corresponds to a sequence extension $\langle \pat, \trfevent_{\node'} \rangle$, we must have $\event_{\node''} = \event_{\node'}, \itmset_{\node''} > \itmset_{\node}$. This contradicts $\itmset_{\ancest_{\node'}} \le \itmset_{\node}$, by definition of ancestor label $\ancest_{\node'}$.

    For an itemset extension $\langle \pat, \event_{\node'} \rangle$, assume by contradiction that $\tpath = (\node, \dots, \node')$ is minimal, but, 
    \begin{enumerate}
        \item $\itmset_{\node'} \neq \itmset_{\node}$ and \label{cond1}
        \item $\exists \event_j \in \itmset^{|\pat|}: \nexists \node'' \in \tpath: \event_{\node''} = \event_j, \itmset_{\node''} = \itmset_{\node'}$, or \label{cond2}
        \item $\forall \event_j \in \itmset^{|\pat|}, \exists \node'' \in \tpath: \event_{\node''} = \event_j, \itmset_{\node''} = \itmset_{\node'}$ and \label{cond3}
        \begin{enumerate}
            \item $\itmset_{\ancest_{\node'}} \ge \itmset_{\node}$, and \label{cond4}
            \item $\exists \node''' \in \tpath: \event_{\node'''} = \event_{\node'}$ and $\node''' \in \nodeset\left(\langle \pat, \event_{\node'} \rangle\right)$. \label{cond5}
        \end{enumerate}
    \end{enumerate}
    Consider conditions~\ref{cond1} and \ref{cond2}. Due to condition~\ref{cond1}, node $\node'$ models an event within another itemset to the event modeled by node $\node$. As path $\tpath$ is minimal, we must have $\itmset^{|\pat|} \sqsubset \itmset^{\itmset_{\node'}}$. This means $\forall \event_j \in \itmset^{|\pat|}, \exists \node'' \in \tpath: \event_{\node''} = \event_j, \itmset_{\node''} = \itmset_{\node'}$, a contradiction to condition~\ref{cond2}. Consider conditions~\ref{cond1} and \ref{cond3}. Due to condition \ref{cond5}, path $(\node, \dots, \node'')$ is minimal, a contradiction to the minimality of path $\tpath$.

    For the converse, assume by contradiction that the statement holds, but path $\tpath$ is not minimal. Then there must exist a node $\node''' \in \tpath, \node''' \neq \node'$ such that path $(\node, \dots, \node'')$ is minimal for the construction of set $\nodeset\left(\langle \pat, \event_{\node'} \rangle\right)$. This contradicts $\itmset_{\ancest_{\node'}} < \itmset_{\node}$ by the definition of ancestor label $\ancest_{\node'}$, and contradicts $\node''' \not\in \nodeset\left(\langle \pat, \event_{\node'} \rangle\right)$ otherwise.
\end{proof}

At each iteration, $\btrie$Miner takes a tuple $(\pat, \nodeset(\pat))$ and extends $\pat$ by searching the sub-tries rooted at nodes $\node \in \nodeset(\pat)$ and checking the conditions of Theorem~\ref{thm:ancestor}. The process can be made more efficient by following a two-phase depth-first-search procedure: 
\begin{enumerate}[label={\upshape\bfseries Phase \arabic*:},wide = 0pt]
\item Search the paths rooted at node $\node$ up until the first node $\node': \itmset_{\node'} \neq \itmset_\node$. Here, the conditions of Theorem~\ref{thm:ancestor} are automatically satisfied, with nodes $\node'' \neq \node'$ modeling an itemset extension and nodes $\node'$ modeling a sequence extension, as proved in Lemma~\ref{lem:twophase}.
\begin{lemma} \label{lem:twophase}
    Let $\tpath = (\node, \dots, \node') \in \btrie$ be a path rooted at a node $\node \in \nodeset(\pat)$ and terminating at the first node $\node': \itmset_{\node'} \neq \itmset_{\node}$. The node $\node'$ models a sequence extension $\langle \pat, \trfevent_{\node'} \rangle$, and all nodes $\node'' \in \tpath: \node'' \neq \node, \node'' \neq \node'$ model an itemset extension $\langle \pat, \event_{\node''} \rangle$.
\end{lemma}
\begin{proof}
    By the statement, all nodes $\node'' \in \tpath: \node'' \neq \node'$ have $\itmset_{\node''} = \itmset_{\node}$ and thus belong to the same itemset. By definition, events within an itemset are unique and cannot be repeated. Therefore, we have $\itmset_{\ancest_{\node''}} < \itmset_{\node}$, which by Theorem~\ref{thm:ancestor}, means that all nodes $\node''$ model an itemset extension $\langle \pat, \event_{\node''} \rangle$. 

    Similarly, as node $\node'$ is the first node in $\tpath$ such that $\itmset_{\node'} \neq \itmset_{\node}$, its ancestor $\ancest_{\node'}$ must either occur prior to node $\node$ or within the same itemset as node $\node$. This means $\itmset_{\ancest_{\node'}} \le \itmset_{\node}$, which by Theorem~\ref{thm:ancestor}, means that node $\node'$ models a sequence extension $\langle \pat, \event_{\node'} \rangle$. 
\end{proof}

\item Search the paths rooted at nodes $\node'$ found in phase 1, and check the conditions of Theorem~\ref{thm:ancestor} to determine itemset and sequence extensions for any traversed node. 
\end{enumerate}
The complete $\btrie$Miner algorithm for $\btrie$ is given in Algorithm~\ref{alg:btrieminer}, proved to find all frequent patterns in Theorem~\ref{thm:soundalgo}, and exemplified in Example~\ref{ex:SPM}.
\begin{algorithm}
	\caption{The $\btrie$Miner algorithm.} \label{alg:btrieminer}
	\begin{algorithmic}[1]
        \State Initialize pattern-node pairs $\left\{  \left( \pat, \nodeset(\pat) \right)\right\}$ for all $\pat = \langle \trfevent \rangle, \; \event \in \eventspace: \supp(\event) \ge \minfreq$.
        \For{each pair $(\pat, \nodeset(\pat))$}
            \State Let $\nodeset\left(\langle \pat, \event\rangle\right) = \emptyset$ for all $\event \in \trfeventspace$, and $\bar{j} = \max_{j' \in \lbrace 1,\dots,|\pat|-1\rbrace} j' : \itmset_{j'} \neq \itmset_{j'+1}$.
            \For{each $\node \in \nodeset(\pat)$}
                \State Initialize DFS queues $Q^1_{\nodeset} = \langle\node\rangle$, $Q^2_{\nodeset} = \langle\rangle$, $Q_{\itmset} = \langle\rangle$.
                \While{$Q^1_\nodeset$ is nonempty}
                    \State Take $\node'$ from the front of queue $Q_\nodeset$.
                    \For{each child (or child-sibling) node $\node''$ of node $\node'$}
                        \If{$\itmset_{\node'} = \itmset_{\node''}$}
                            \State Add node $\node''$ to $\nodeset\left(\langle \pat, \event_{\node''}\rangle\right)$ as an itemset extension,
                            \State Add node $\node''$ to the front of queue $Q^1_\nodeset$.
                        \Else
                            \State Add node $\node''$ to $\nodeset\left(\langle \pat, \trfevent_{\node''}\rangle\right)$ as a sequence extension,
                            \State Add node $\node''$ to the front of queue $Q^2_\nodeset$.
                            \If{$\event_{\node''} = \event_{\bar{j}}$}
                            \State Add 1 to the front of queue $Q_\itmset$.
                            \Else
                                \State Add 0 to the front of queue $Q_\itmset$.
                            \EndIf
                        \EndIf
                    \EndFor
                \EndWhile
                \While{$Q^2_\nodeset$ is nonempty}
                    \State Take $\node'$ from the front of queue $Q^2_\nodeset$, and integer $k$ from front of queue $Q_\itmset$.
                    \For{each child (or child-sibling) node $\node''$ of node $\node'$}
                        \If{$\itmset_{\node'} = \itmset_{\node''}$}
                            \If{$k = |\pat| - j$ and conditions of Theorem~\ref{thm:ancestor} are satisfied}
                                \State Add node $\node''$ to $\nodeset\left(\langle \pat, \event_{\node''}\rangle\right)$ as an itemset extension.
                            \EndIf
                            \If{Conditions of Theorem~\ref{thm:ancestor} are satisfied}
                                \State Add node $\node''$ to $\nodeset\left(\langle \pat, \trfevent_{\node''}\rangle\right)$ as a sequence extension.
                            \EndIf
                            \State Add node $\node''$ to the front of queue $Q^2_\nodeset$.
                            \If{$k < |\pat| - j$ and $\event_{\node''} = \event_{\bar{j} + k}$}
                                \State Add k + 1 to the front of queue $Q_\itmset$.
                            \Else
                                \State Add k to the front of queue $Q_\itmset$.
                            \EndIf
                        \Else
                            \If{conditions of Theorem~\ref{thm:ancestor} are satisfied}
                                \State Add node $\node''$ to $\nodeset\left(\langle \pat, \trfevent_{\node}\rangle\right)$ as a sequence extension.
                            \EndIf
                            \State Add node $\node''$ to the front of queue $Q^2_\nodeset$.
                            \If{$\event_{\node''} = \event_{\bar{j}}$}
                                \State Add 1 to the front of queue $Q_\itmset$.
                            \Else
                                \State Add 0 to the front of queue $Q_\itmset$.
                            \EndIf
                        \EndIf
                    \EndFor
                \EndWhile
            \EndFor 
            \For{each $\event \in \trfeventspace$}
                \If{$\sum_{\node \in \nodeset\left(\langle \pat, \event\rangle\right)}\freq_\node \ge \minfreq$}
                    \State Add $\left(\langle \pat, \event \rangle, \nodeset\left(\langle \pat, \event\rangle\right)\right)$ to the set of pattern-node pairs.
                \EndIf
            \EndFor
        \EndFor
        \State \textbf{return} Set of frequent patterns $\pat: \sum_{\node \in \nodeset\left(\pat\right)}\freq_\node \ge \minfreq$. 
	\end{algorithmic}
\end{algorithm}
\begin{theorem}\label{thm:soundalgo}
    A pattern is frequent if and only if it is found by $\btrie$Miner.
\end{theorem}
\begin{proof}
    Assume that a pattern $\pat$ is frequent, and thus has support $\supp(\pat) \ge \minfreq$. By the definition of support values, there exists at least $\minfreq$ sequences $\seq_i \in \seqdata$ such that $\pat \sqsubseteq \seq_i$. By Proposition~\ref{prop:minprefix}, node set $\nodeset(\pat)$ models all minimum-sized prefixes $\prefix_i \sqsubseteq \seq_i: \pat \sqsubseteq \prefix_i$ for all $\seq_i \in \seqdata$. By Proposition~\ref{prop:mine} and Theorem~\ref{thm:ancestor}, all nodes belonging to set $\nodeset(\pat)$ are found by $\btrie$Miner. Therefore, $\btrie$Miner finds all sequences $\seq_i: \pat \sqsubseteq \seq_i$.

    Assume that a pattern $\pat$ is found by $\btrie$Miner. This means that $\sum_{\node \in \nodeset(\pat)} \freq_\node \ge \minfreq$. As nodes $\node \in \nodeset(\pat)$ model a minimum-sized prefix $\prefix_i \sqsubseteq \seq_i: \pat \sqsubseteq \prefix_i$, then there must exist at least $\minfreq$ sequences $\seq_i \in \seqdata$ such that $\pat \sqsubseteq \seq_i$. Therefore, $\supp(\pat) \ge \minfreq$, and pattern $\pat$ is frequent.
\end{proof}
\begin{example}\label{ex:SPM}
    \sloppy In the first iteration, $\btrie$Miner initializes with pattern-node pairs $(\langle\bar{a}\rangle, \lbrace\node_1\rbrace), (\langle\bar{b}\rangle, \lbrace\node_2, \node_7, \node_{16}\rbrace), (\langle\bar{c}\rangle, \lbrace\node_5, \node_8, \node_{11}, \node_{15}\rbrace)$. The mining algorithm takes a pattern-node pair, for example, $(\langle\bar{a}\rangle, \lbrace\node_1\rbrace)$, and in the first phase performs a depth-first-search of the subtrie rooted at node $\node_1$ to find the first nodes $\node': \itmset_{\node'} \neq \itmset_\node$. This results in finding nodes $\node_3, \node_6, \node_7, \node_9$. All these nodes model a sequence extension according to Lemma~\ref{lem:twophase}, and give a pattern-node pair $\left(\langle\bar{a}, \bar{a}\rangle, \lbrace\node_3, \node_9\rbrace), (\langle\bar{a}, \bar{b}\rangle, \lbrace\node_7\rbrace), (\langle\bar{a}, \bar{c}\rangle, \lbrace\node_6\rbrace)\right)$. On the other hand, any node traversed on the paths $\tpath = (\node_1, \dots, \node'), \node' \in \{\node_3, \node_6, \node_7, \node_9\}$ models an itemset extension according to Lemma~\ref{lem:twophase}, which gives pattern-node pairs $(\langle\bar{a}, b\rangle, \lbrace\node_2\rbrace) (\langle\bar{a}, c\rangle, \lbrace\node_5, \node_8\rbrace)$.
    
    In the second phase, search is initiated from nodes $\node_3, \node_6, \node_7, \node_9$, found in phase 1. Any pattern extension is determined by following Theorem~\ref{thm:ancestor}. For example, Searching the sub-trie rooted at node $\node_9$ first traverses node $\node_{10}$. To check for a sequence extension, we have $\itmset_{\node_{10}} \neq \itmset_{\node_1}$ and $\itmset_{\ancest_{\node_{10}}} \le \itmset_{\node_1}$, which satisfies the condition of Theorem~\ref{thm:ancestor}. The corresponding pattern-node pair is then updated to $\left(\langle\bar{a}, \bar{c}\rangle, \lbrace\node_6, \node_{10}\rbrace)\right)$. 
    
    To check for an itemset extension, we have $\itmset_{\node_{10}} \neq \itmset_{\node_1}$ which violates the first condition of Theorem~\ref{thm:ancestor}, but $\event_{\node_{9}} = \event_{\node1}, \itmset_{\node_9} = \itmset_{\node_{10}}$ which satisfies the first part of the second condition. Checking the second parts of the second condition, we have $\itmset_{\ancest_{\node_{10}}} = \itmset_{\node_1}$, and $\node_8 \in \nodeset(\langle \bar{a}, c \rangle)$ which violates both conditions. Therefore, node $\node_{10}$ does not model an itemset extension.
\end{example}
The worst-case space complexity of $\btrie$Miner is given by Theorem~\ref{thm:spacebtrie}.
\begin{theorem}\label{thm:spacebtrie}
    The worst-case space complexity of $\btrie$Miner is $\order\left(\min\lbrace \numseq, \numevent^{\maxseqsize}\rbrace\maxseqsize\log(\numseq) \right)$, and \newline $\order\left(\min\lbrace \numseq, \numevent^{\maxseqsize}\rbrace\maxseqsize \right)$ for reasonably-sized integers $\numseq$.
\end{theorem}
\begin{proof}
    A node $\node \in \btrie$ stores integer values $\freq_\node$, $\event_\node$, $\itmset_\node$, a positional value $\ancest_\node$, and two positional values $\child(\node)$, $\sibl(\node)$. All values are bounded by $\order(\log(\numseq))$.

    As trie models of the data set are identical in their node set $\nodeset$, and by Lemma~\ref{lem:spacetminer}, the number of nodes in a trie model is $\order\left(\min\lbrace \numseq, \numevent^{\maxseqsize}\rbrace\maxseqsize\right)$. This gives the total worst-case memory complexity of $\btrie$ as $\order\left(\min\lbrace \numseq, \numevent^{\maxseqsize}\rbrace\maxseqsize\log(\numseq)\right)$. 

    Similarly to TreeMiner, $\btrie$ Miner stores a positional integer $j \le \maxseqsize$ for at most every node $\node \in \btrie$, bounding its worst-case space complexity by $\order\left(\min\lbrace \numseq, \numevent^{\maxseqsize}\rbrace\maxseqsize\log(\numseq)\right)$. The total space complexity of $\btrie$Miner is thus $\order\left(\min\lbrace \numseq, \numevent^{\maxseqsize}\rbrace\maxseqsize\log(\numseq) \right)$. Reasonably sized integers $\numseq$ consume constant memory and reduce the complexity to $\order\left(\min\lbrace \numseq, \numevent^{\maxseqsize}\rbrace\maxseqsize \right)$.
\end{proof}
Compared to TreeMiner, $\btrie$Miner is at least $\order(\numevent^2)$ times more memory efficient and never larger. Similarly, $\btrie$Miner is always asymptotically smaller than a vector-based SPM algorithm. In the best case for $\btrie$Miner, we have $\numevent = 1$ and space complexity of $\maxseqsize\log(\numseq)$, leading to $\order(\numseq)$ times more efficient memory consumption than a vector model for reasonably-sized integers $\numseq$. In the worst case, $\btrie$ uses the same number of nodes as entries $\sum_{\seq_i \in \seqdata} |\seq_i|$. In practice, this leads to higher memory usage due to the memory overhead of label sets $\labset_\node$ stored at each node $\node \in \btrie$. We address this deficiency by developing a hybrid model in the next section. 

\subsection{Hybrid Tries and the $\htrie$Miner Algorithm} \label{sec:htrie}

Trie models are most memory-efficient when modeling sequences $\seq_i \in \seqdata$ that highly overlap. Such sequences can be represented by fewer nodes $|\nodeset|$, leading to a more compact model that compensates for the higher memory overhead $\memtri$ of each trie node. On the other hand, when many sequences $\seq_i \in  \seqdata$ do not overlap, the model is less compact and can lead to increased memory consumption in practice. For such data sets, vector models are a more memory-efficient approach. 

Regardless of the overlap for sequences $\seq_i \in \seqdata$, sequential data sets generally have a high overlap on their prefixes $\prefix_i: \prefix_i \sqsubseteq \seq_i \in \seqdata$. This is due to the lower number of possible event combinations $|\eventspace|^{j} \ge |\nodeset_j|$ at any length $j \le |\seq_i|$, which is typically much lower than $\numseq$ for smaller values of $j$. For such prefixes, a trie model is more memory efficient. As $j$ increases, so does the possible number of unique and non-overlapping sequences, leading to potentially less overlap on postfixes $\postfix_i \sqsubseteq \seq_i \in \seqdata$. For such postfixes, a vector model is more memory efficient in practice. We propose a hybrid model $\htrie$ to take advantage of this trade-off. 

\begin{figure}
    \centering
    \includegraphics[width=0.7\textwidth]{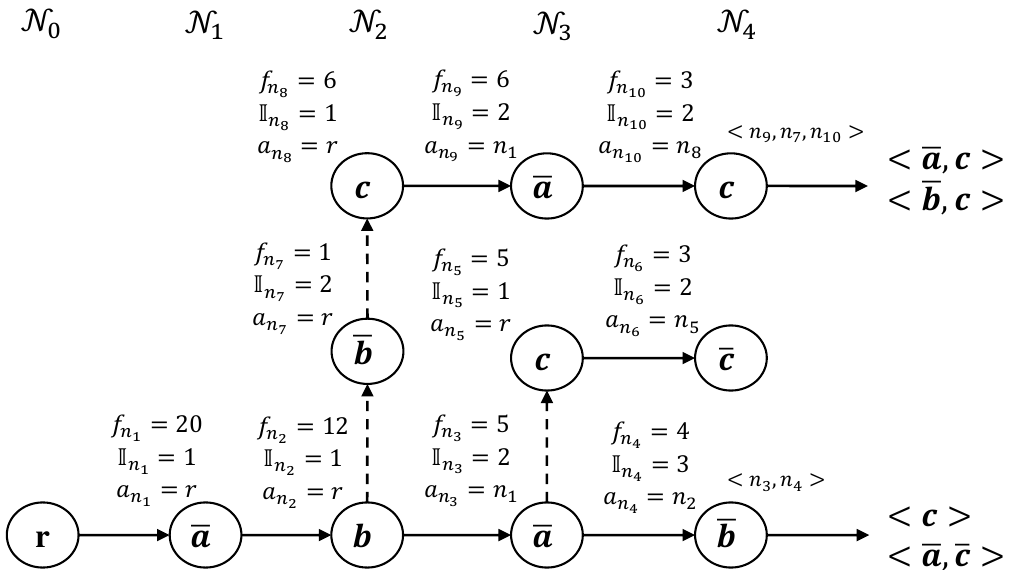}
        \captionsetup{justification=centering}
    \caption{An $\htrie$ model of the data set in Table~\ref{tab:vector}. The model transitions from a trie to a vector representation at layer $\nodeset_4$.  Transitioning nodes $\node_4, \node_{10}$ are associated to an ancestor vector that tracks the ancestors for all events $\event \in \eventspace: \event \in \transformb(\rootnode, \dots, \node_j), j \in \{4,10\}$.}
    \label{fig:htrie}
\end{figure}

A hybrid model $\htrie$ is a $\btrie$ representation of all prefixes $\prefix_i \sqsubseteq \seq_i$ of length $|\prefix_i| = \optlen$, that transitions into a vector model for the remaining postfixes $\postfix_i: |\postfix_i| = |\seq_i| - \optlen$. For the subsequent mining task, $\htrie$ also stores at each node $\node$ of the transitioning layer $\nodeset_{\optlen}$, a vector of ancestor nodes $\vec{\ancest} = \langle\ancest_{\event_1}, \dots, \ancest_{\event_{|\vec{\ancest}|}} \rangle$ such that $\ancest_{\event} \in \vec{\ancest}$ if $\event \in \transformb\left(\rootnode, \dots, \node \right)$. For example, Figure~\ref{fig:htrie} illustrates an $\htrie$ model for our running example that transitions at length $\optlen = 4$.

An important challenge of our hybrid model is determining the transition length $\optlen$ that provides the highest memory efficiency. This is data set dependent and determined based on the trade-off between the compression that a $\btrie$ model can provide for layers $\layer_j, 0 < j \le \optlen$, versus the memory overhead introduced by the transition to a vector model for postfixes $\postfix_i: |\postfix_i| = \maxseqsize - \optlen$. We choose $\optlen$ using Theorem~\ref{thm:transition}, where $\numseq_j$ denotes the number of sequences $\seq_i \in \seqdata: |\seq_i| \ge j$, recorded during the preprocessing step of constructing $\btrie$.
\begin{theorem}\label{thm:transition}
    It is most memory-efficient for a hybrid model $\htrie$ to transition from a $\btrie$ representation to a vector representation at the layer $\nodeset_\optlen$, where \newline \small$\optlen = \underset{\optlen \in \{0, \dots, \maxseqsize\}}{\arg\min} \memtri\memint\sum\limits_{j = 1}^{\optlen}|\nodeset_j| + \mathds{1}(\optlen < \maxseqsize)\left(\memvec + \memint\min\{\numevent, \optlen\}\right)|\nodeset_{\optlen}| + \memvec\numseq_{\optlen + 1} + \memint\sum\limits_{j = \optlen + 1}^{\maxseqsize} \numseq_j.$\normalsize 
\end{theorem}
\begin{proof}
    For an $\htrie$ transitioning at a layer $\nodeset_{j'}\in \htrie: j' \in \{0, \dots, \maxseqsize\}$ involves a $\btrie$ model of all prefixes $\prefix_i \sqsubseteq \seq_i \in \seqdata: |\prefix_i| = j'$, and a vector model of postfixes $\postfix_i \sqsubset \seq_i: |\postfix_i| = |\seq_i| - j'$. 

    The $\btrie$ model of prefixes $\prefix_i$ consumes $\memtri\memint$ memory overhead per nodes $\node \in \btrie$. If $j' < \maxseqsize$, $\htrie$ also stores a vector of at most $\min\{\numevent, \optlen\}$ ancestors per node $\node \in \nodeset_{j'}$, which has $\memvec + \memint\min\{\numevent, j'\}$ memory overhead. The total memory consumption of the $\btrie$ model of $\htrie$ is thus $\memtri\memint\sum_{j = 1}^{j'} |\nodeset_j| + \mathds{1}(\optlen < \maxseqsize)\left(\memvec + \memint\min\{\numevent, j'\}\right)|\nodeset_{j'}|$.

    The vector model of postfixes $\postfix_i$ uses one vector per $\numseq_{j' + 1}$ sequences $\seq_i \in \seqdata: |\seq_i| \ge j' +1$, and $\memint$ memory overhead per events $\event \in \postfix_i$. This gives the total memory consumption of the vector model of $\htrie$ as $\memvec\numseq_{j' + 1} + \memint\sum\limits_{j = j' + 1}^{\maxseqsize} \numseq_j$.

    The most memory-efficient transition layer $\nodeset_{\optlen}$ is determined by the $\htrie$ model with the lowest memory consumption, that is, \newline\small$\optlen = \underset{\optlen \in \{0, \dots, \maxseqsize\}}{\arg\min} \memtri\memint\sum\limits_{j = 1}^{\optlen}|\nodeset_j| + \mathds{1}(\optlen < \maxseqsize)\left(\memvec + \memint\min\{\numevent, \optlen\}\right)|\nodeset_{\optlen}| + \memvec\numseq_{\optlen + 1} + \memint\sum\limits_{j = \optlen + 1}^{\maxseqsize} \numseq_j.$\normalsize
\end{proof}
\begin{algorithm}
	\caption{Construction of $\htrie$.} \label{alg:htrie}
	\begin{algorithmic}[1]
        \State Filter the data set and remove all events $\event \in \eventspace: \supp(\event) < \minfreq$, and record $\numseq_j$, $\numevent_j$ for all $j \in \lbrace 0, \dots, \maxseqsize \rbrace$.
        \State Determine $\optlen$ according to Theorem~\ref{thm:transition} and values $\numseq_j$ and $\numevent_j$.
        \For{each $\seq_i \in \seqdata$}
            \State Construct a $\btrie$ model for prefix $\prefix_i \sqsubseteq \seq_i: |\prefix_i| = \optlen$.
            \State Add child arc $(\node, \node')$ to $\arcset$, where $\node$ is the last node constructed by $\btrie$ and $\node'$ is contains the vector that models $\postfix_i: \seq_i = \langle \prefix_i, \postfix_i\rangle$. 
            \State Store ancestor vector constructed at $\btrie$ at $\node'$.
        \EndFor
		\State \textbf{return} $\htrie$. 
	\end{algorithmic}
\end{algorithm}

Theorem~\ref{thm:transition} is based on the size of each layer $\nodeset_{j} \in \btrie$. Unfortunately, these values are unknown prior to the construction of the $\btrie$ model, but following Theorem~\ref{thm:spacebtrie}, can be approximated by the upper bound $|\nodeset_{j}| \le  \min\{\prod_{j' = 0}^j\numevent_{j'}, \numseq_j\}$. Here, $\numevent_j$ is the number of unique events at the length $j$ of sequences $\seq_i \in \seqdata$, that are recorded during preprocessing. Length $\optlen$ is then found by iterating over $j \in \lbrace 0,\dots,\maxseqsize \rbrace$ and calculating the equation of Theorem~\ref{thm:transition}. In the extreme case a trie representation does not provide any memory saving potential, we have $\optlen = 0$ and $\htrie = \vecmodel$. On the other hand, if a trie model of the entire data set is predicted to be more efficient, we have $\optlen = \maxseqsize$ and $\htrie = \btrie$. Algorithm~\ref{alg:htrie} gives the full construction procedure for $\htrie$, and Example~\ref{ex:htrie} demonstrates the procedure of determining $\optlen$ for our running example.
\begin{example}\label{ex:htrie}
    We have $\numseq_j = \langle 20, 19, 16, 10, 4, 3 \rangle$, and $\numevent_j = \langle 1, 3, 2, 3, 3, 2 \rangle$. For our system, we have $\memint = 4, \memvec = 24$, and by Theorem~\ref{thm:spacebtrie}, $\memtri = 6$. Checking the equation of Theorem~\ref{thm:transition}, we have the following memory estimates for $j \in \{0,\dots,6\}$:
    \small
    \begin{enumerate}[label={\upshape $j = $ \arabic*:},wide = 0pt]
        \itemsep0em
        \setcounter{enumi}{-1}
        \item $24\times20 + 4\times(20 + 19 + 16 + 10 + 4 + 3) = 768.$
        \item $6\times4\times1 + (24 + 4\times1)\times1 + 24\times 19 + 4\times(19 + 16 + 10 + 4 + 3) = 716.$
        \item $6\times4\times(1+3) + (24 + 4\times2)\times3 + 24\times16+ 4\times(16 + 10 + 4 + 3) = 708.$
        \item $6\times4\times(1+3+6) + (24 + 4\times3)\times6 + 24\times10+ 4\times(10 + 4 + 3) = 764.$
        \item $6\times4\times(1+3+6+10) + (24 + 4\times3)\times10 + 24\times4+ 4\times(4 + 3) =964.$
        \item $6\times4\times(1+3+6+10+4) + (24 + 4\times3)\times4 + 24\times3 + 4\times(3) =804.$
        \item $6\times4\times(1+3+6+10+4+3)=648.$
    \end{enumerate}\normalsize
    We thus have $\optlen = \maxseqsize = 6$, and $\htrie = \btrie$ for our running example.
\end{example}

\begin{algorithm}[t!]
	\caption{The $\htrie$Miner algorithm.} \label{alg:htrieminer}
	\begin{algorithmic}[1]
        \State Initialize pattern-node pairs $\left\{  \left( \pat, \nodeset(\pat) \right)\right\}$ for all $\pat = \langle \trfevent \rangle, \; \event \in \eventspace, \event \in \prefix_i: |\prefix_i| \le \optlen, \supp(\event) \ge \minfreq$.
        \State Initialize pattern-position pairs $\left\{ \left(\pat, (i,j)\right) \right\}$ for all $\pat = \langle \trfevent \rangle, \; \event \in \eventspace, \event \in \postfix_i: |\postfix_i| = |\seq_i| - \optlen, \seq_i \in \seqdata, \supp(\event) \ge \minfreq$.
        \For{each pair $(\pat, (i,j))$}
            \State Mine the vector $\vecmodel$ pointed to by $i$ starting from position $j$ according to prefix-projection.
            \State Add pattern-position pairs $\left\{ \left(\langle \pat, \event_{j'} \rangle, (i,j')\right) \right\}$ for all events $\event_j \in \vecmodel$ that give an extension for $\pat$. 
        \EndFor
        \For{each pair $(\pat, \nodeset(\pat))$}
            \For{all nodes $\node \in \nodeset(\pat)$}
                \State Traverse all paths $\tpath = (\node, \dots, \node') \in \btrie$ part of $\htrie$ and mine it according to Algorithm~\ref{alg:btrieminer}.
                \For{all vector children $\vecmodel$ pointed by node $\node' \in Q^1_\nodeset$ in $\btrie$Miner}
                    \State Mine the vector $\vecmodel$ starting from position $1$ according to prefix-projection.
                    \State Add pattern-position pairs $\left\{ \left(\langle \pat, \event_{j} \rangle, (i,j)\right) \right\}$ for all events $\event_j \in \vecmodel$ that give an extension for $\pat$ using ancestor vector $\vec{\ancest}$, and according to Theorem~\ref{thm:ancestor}.
                \EndFor
                \For{all vector children $\vecmodel$ pointed by node $\node' \in Q^2_\nodeset$ in $\btrie$Miner}
                    \State Mine the vector $\vecmodel$ starting from position $1$ according to prefix-projection.
                    \State Add pattern-position pairs $\left\{ \left(\langle \pat, \event_{j} \rangle, (i,j)\right) \right\}$ for all events $\event_j \in \vecmodel$ that give an extension for $\pat$ according to Theorem~\ref{thm:ancestor}.
                \EndFor
            \EndFor 
        \EndFor
        \State \textbf{return} Set of frequent patterns $\pat: \sum_{\node \in \nodeset\left(\pat\right)}\freq_\node \ge \minfreq$. 
	\end{algorithmic}
\end{algorithm}

To mine $\htrie$ models, we combine $\btrie$Miner and prefix-projection into a novel $\htrie$Miner algorithm. The $\htrie$Miner algorithm mines its trie model of prefixes $\prefix_i$ using the procedure of $\btrie$Miner, and its vector model of postfixes $\postfix_i$ using prefix-projection. If the mining algorithm traverses a transitioning node $\node \in \nodeset_{\optlen}$, it uses the ancestor vector $\vec{\ancest}$ and Theorem~\ref{thm:ancestor} to determine valid pattern extensions in the vector model. The complete procedure is given in Algorithm~\ref{alg:htrieminer}.

The worst-case space complexity of $\htrie$Miner is given by Lemma~\ref{lem:spacehybrid}.
\begin{lemma} \label{lem:spacehybrid}
    The worst-case space complexity of $\htrie$Miner is $\order\left(\numseq_{\optlen}\maxseqsize\log(\numseq)\right)$, and \newline $\order\left(\numseq_{\optlen}\maxseqsize\right)$ for reasonably-sized integers $\numseq_{\optlen}$. 
\end{lemma}
\begin{proof}
    The largest layer of $\htrie$ is $\nodeset_{\optlen}$, which is bounded by $|\nodeset_{\optlen}| \le \min\{\numevent^{\optlen}, \numseq_{\optlen}\}$. At each node of the transitioning layer $\nodeset_{\optlen}$ the ancestor vector uses $\order(\min\{\numevent^{\optlen}, \numseq_{\optlen}\}\min\{\numevent, \optlen\})$ memory. By Theorem~\ref{thm:spacebtrie} this gives the worst-case space complexity of the $\btrie$ part of $\htrie$ and its corresponding mining algorithm as $\order\left(\min\{\numevent^{\optlen}, \numseq_{\optlen}\}\optlen\log(\numseq)\right)$.

    The largest number of sequences of the vector model of $\htrie$ is $\numseq_{\optlen + 1}$. By Lemma~\ref{lem:vecspace}, this has a worst-case space complexity of $\order\left(\numseq_{\optlen + 1}(\maxseqsize - \optlen)\log(\numseq)\right)$. As the number of sequences are non-increasing in sequence length, we have $\numseq_{\optlen +1} \le \numseq_{\optlen}$. This gives the worst-case space complexity of the vector part of $\htrie$ and its corresponding algorithm as $\order\left(\numseq_{\optlen}(\maxseqsize - \optlen)\log(\numseq)\right)$.

    Adding the two complexities together gives the total worst-case space complexity of \newline $\order\left(\min\{\numevent^{\optlen}, \numseq_{\optlen}\}\optlen\log(\numseq) + \numseq_{\optlen}(\maxseqsize - \optlen)\log(\numseq)\right) \le \order(\numseq_{\optlen}\maxseqsize\log(\numseq))$. Reasonably sized integers $\numseq$ consume constant memory and reduce the complexity to $\order(\numseq_{\optlen}\maxseqsize)$.
\end{proof}

As $\optlen \le \maxseqsize$, $\htrie$Miner is always asymptotically smaller and more memory efficient than both $\btrie$Miner and vector-based SPM algorithms. Moreover, by Theorem~\ref{thm:transition}, $\htrie$Miner is also more memory efficient than $\btrie$Miner and vector-based SPM in practice. We indeed observe this in our numerical results, given in the following section. 

\section{Numerical Results}\label{sec:results}

For our numerical tests, we evaluated the effect of different data set models on the performance of state-of-the-art mining algorithms in large-scale SPM. We use PrefixSpan by \citet{han2001prefixspan}, which forms the basis of almost all state-of-the-art vector-based prefix-projection algorithms, and TreeMiner \citep{rizvee2020tree}, which is the only available trie model for SPM. Note that TreeMiner uses several additional mining techniques, such as co-occurrence information of events \citep{fournier2014fast} for faster SPM, which are not implemented in our mining algorithms for a base-case comparison. Nonetheless, any such algorithmic enhancement developed for time efficiency in the rich literature of SPM can also be implemented on our models without loss of generality. 

All algorithms were coded in C++ and executed on a PC with an Intel Xeon W-2255 processor, 256GB of memory, and Ubuntu 20.04.1 operating system.\footnote{Our algorithms are open source and available at \href{https://github.com/aminhn/BDTrie}{https://github.com/aminhn/BDTrie}} Note that our system memory is considerably higher than an average PC, often ranging between 8-32GB. We limit all tests to one core of the CPU and a 36,000-second time limit. 

\subsection{Data Sets}\label{sec:resultsdata}

\begin{table}
\centering
 \caption{Data sets.}
\label{tab:datasets}
        \resizebox{\textwidth}{!}{
	\begin{tabular}{l l c c c c c c c}\toprule
	Size & Name & $\numseq$ & $\numevent$ & $\maxseqsize$& $avg(|\seq_i|)$ & $max(|\itmset^j|)$ & $avg(|\itmset^j|)$ & $\sum_{\seq_i \in \seqdata} |\seq_i|$ \\\midrule
        \multirow{2}{*}{Large} 
        & Criteo & 4,373,472,329 & 214 & 20 & 20 & 1 & 1 & 87,469,446,580 \\
        & Genome & 2,049,780,092 & 20 & 37 & 13 & 1 & 1& 26,485,605,043 \\ \midrule
        \multirow{2}{*}{Medium}
	& Twitch & 15,524,308 & 790,100 & 456 & 30.3 & 170 & 1.2 & 469,655,703 \\
	& Spotify &  124,950,054 & 5 & 20 & 2.4 & 5 & 1.1 &  294,302,842\\ \midrule
        \multirow{2}{*}{Small}
	& Kosarak &  990,002 & 41,270 & 2,498 & 8.1 & 1 & 1 & 8,019,015\\
	& MSNBC & 989,818 & 17 & 14,795 & 4.8 & 1 & 1 &  4,698,794\\ \bottomrule
	\end{tabular}}
\end{table}

\begin{table}
\centering
 \caption{Number of frequent patterns by data set and support threshold.}
\label{tab:numpatt}
        \resizebox{0.7\textwidth}{!}{
	\begin{tabular}{c c c c c c c}\toprule
	  \multicolumn{7}{c}{Criteo (large)}\\\midrule
        $\minthr$ & 10\% & 8\% & 6\% & 4\% & 2\% & 0.5\% \\ \cmidrule{1-7}
        Patterns found & 10,705 & 20,132 & (41,559) & (78,598) & (88,434) & (182,541)  \\ \bottomrule
        \multicolumn{7}{c}{Genome (large)}\\\midrule
        $\minthr$ & 10\% & 8\% & 6\% & 4\% & 2\% & 0.5\% \\ \cmidrule{1-7}
        Patterns found & (264) & (314) & (459) & (552) & (598) & (615) \\ \bottomrule
        \multicolumn{7}{c}{Twitch (medium)}\\\midrule
        $\minthr$ & 8\% & 4\% & 2\% & 1\% & 0.5\% & 0.15\% \\ \cmidrule{1-7}
        Patterns found & 2 & 14 & 160 & 1,382 & 13,493 & 789,685 \\ \bottomrule
        \multicolumn{7}{c}{Spotify (medium)}\\\midrule
        $\minthr$ & 1\% & 1e-1\% & 1e-2\% & 1e-3\% & 1e-4\% & 1e-5\% \\ \cmidrule{1-7}
	  Patterns found & 59 & 377 & 2,559 & 16,253 & 107,257 & 696,098 \\ \bottomrule
        \multicolumn{7}{c}{Kosarak (small)}\\\midrule
        $\minthr$ & 4\% & 2\% & 1\% & 0.5\% & 0.25\% & 0.1\% \\ \cmidrule{1-7}
        Patterns found & 29 & 94 & 329 & 1,462 & 8,427 & 758,141 \\ \bottomrule
        \multicolumn{7}{c}{MSNBC (small)}\\\midrule
        $\minthr$ & 3\% & 1\% & 0.3\% & 0.1\% & 0.03\% & 0.01\% \\ \cmidrule{1-7}
        Patterns found & 22 & 254 & 2,014 & 16,620 & 303,917 & 102,108,060 \\ \bottomrule
        \multicolumn{7}{l}{Note. Parenthesis represent lower bound.}
	\end{tabular}}
\end{table}

We used six real-life data sets in our numerical tests, given in Table~\ref{tab:datasets}. The data sets are chosen based on common applications of SPM, namely, click-stream mining and bioinformatics \citep{fournier2017survey}. The data sets are grouped into three sizes: large, medium and small. 

The small data sets include Kosarak and MSNBC, which have been benchmark click-stream instances for SPM since the early 2010s~\citep{fournier2016spmf}. The medium data sets include the Twitch data set \citep{ rappaz2021recommendation}, and the Spotify data set \citep{brost2019music}. The Twitch data set includes user streaming content on the live streaming platform Twitch. Here, events $\event \in \eventspace$ are defined as a unique streamer, and consecutive streamers watched by a user form a sequence $\seq_i \in \seqdata$. Two consecutive events are assumed to be in the same itemset if they are visited by the user within a one hour time frame. The Spotify data set \citep{brost2019music} is a collection of user music consumption on the media streaming platform Spotify. Events $\event \in \eventspace$ are considered to be the \emph{time-signature} (refer to \citet{brost2019music} for definition of time-signature) of listened to tracks. Events are considered to be in different itemsets if a \emph{context switch} (refer to \citet{brost2019music} for definition of a context switch) occurs between them.

The large data sets include the Criteo 1TB click-stream data set \citep{criteo}, and the amino acid representation of the 1000 Genomes data set \citep{10002015global}. Unfortunately, the full size of both data sets is larger than 1TB in size and impossible to fit into memory for most systems, including ours. We thus use a subset of each data set for our numerical tests. The Criteo data set includes 20 out of the 40 possible events in each sequence, and the Genome data set includes the first subset ``ERR3988796'' of genomes as specified by \citet{amazonaws}.

An important aspect of mining the above data sets is the imposed support threshold $\minthr$. Higher support thresholds lead to fewer frequent patterns, mainly showcasing the time and memory efficiency of building and storing a data set representation in memory. Lower support thresholds lead to a larger number of frequent patterns, mainly showcasing the time and memory efficiency of a mining algorithm. We choose a wide range of thresholds for each data set tailored to observe and showcase both scenarios. The thresholds and number of frequent patterns found for each data set are given in Table~\ref{tab:numpatt}. Note that when all algorithms exceeded the imposed time, the reported number of frequent patterns is a lower bound.

\subsection{Memory Consumption}\label{sec:results:mem}

\begin{figure}
     \centering
    \captionsetup{justification=centering}
     \begin{subfigure}[b]{\textwidth}
         \centering
         \includegraphics[width=0.75\textwidth]{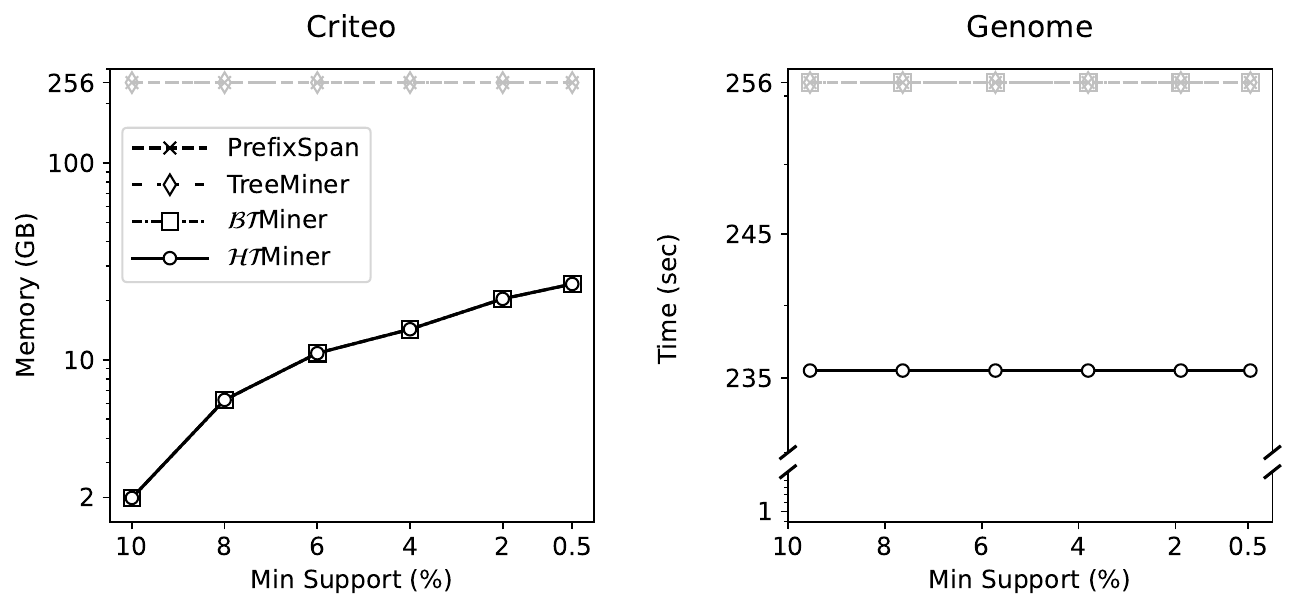}
         \caption{Large data sets. PrefixSpan exceeds system memory at 15\% of the Criteo data set and 62\% of the Genome data set. TreeMiner exceeds system memory during mining of the Criteo data set, and at 3\% of the Genome data set. $\btrie$Miner exceeds system memory at 29\% of the Genome data set. $\htrie$Miner is the only algorithm capable of mining both data sets within system memory. }
     \end{subfigure}
     \quad
     \begin{subfigure}[b]{\textwidth}
         \centering
         \includegraphics[width=0.75\textwidth]{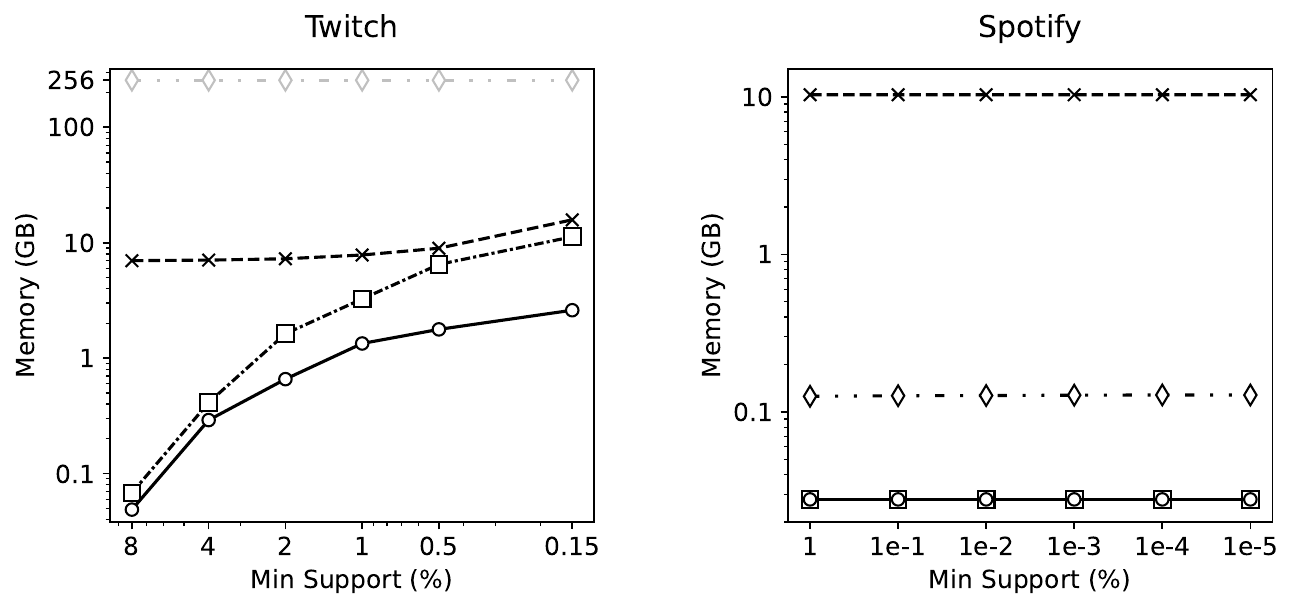}
         \caption{Medium data sets. TreeMiner exceeds system memory at 0.6\% of the Twitch data set.}
     \end{subfigure}
          \quad
     \begin{subfigure}[b]{\textwidth}
         \centering
         \includegraphics[width=0.75\textwidth]{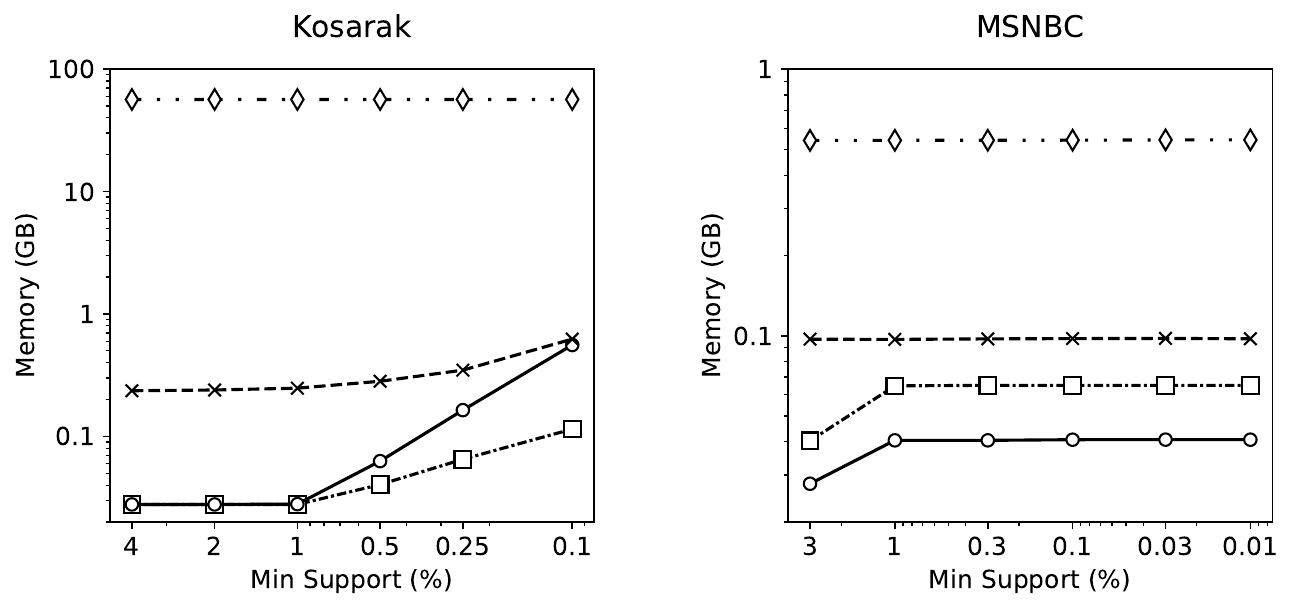}
         \caption{Small data sets.}
     \end{subfigure}
        \caption{Peak memory consumption. Algorithms that exceed system memory are shown in gray.}
        \label{fig:memorycomp}
\end{figure}

Figure~\ref{fig:memorycomp} gives the peak memory consumption of all algorithms. In the small data set Kosarak, all algorithms, with the exception of TreeMiner, consume less than 1GB of memory. TreeMiner uses close to 57GB of memory, mainly due to the higher number of events $\numevent$ in this data set which is detrimental to its memory efficiency. The $\htrie$Miner algorithm uses less than 0.7GB of memory which amounts to an average improvement of 68\% (less than 0.2GB) over PrefixSpan and 99.9\% (more than 56GB) over TreeMiner. In contrast to all other data sets, $\htrie$Miner uses slightly higher memory (0.1GB on average, and 0.4GB at most) compared to $\btrie$Miner in lower support thresholds. This is mainly due to the heuristic selection of the transitioning layer, where values $|\nodeset_j|$ were overestimated, leading to a lower than optimal value $\optlen$.  

In the small data set MSNBC, $\htrie$Miner shows an average improvement of 60\% (less than 0.1GB) over PrefixSpan and 93\% (approximately 0.5GB) over TreeMiner. Similarly, $\btrie$Miner shows an average improvement of 37\% (less than 0.1GB) over PrefixSpan and 89\% (approximately 0.5GB) over TreeMiner. These improvements show the slight advantage of compactly fitting the data in memory using a trie with efficient labels, even for smaller data sets. $\htrie$Miner shows an average improvement of 36\% (less than 0.1GB) over $\btrie$Miner, mainly due to the longer sequences in the MSNBC data set which are more efficiently modeled by a hybrid trie structure.

In the medium data set Twitch, PrefixSpan uses less than 16GB of memory, $\btrie$Miner uses approximately 11GB of memory with an average improvement of 64\% over PrefixSpan, and $\htrie$Miner uses less than 3GB of memory with an average improvement of 89\% over PrefixSpan and 54\% over $\btrie$Miner. The improvement of $\htrie$Miner and $\btrie$Miner is higher for larger support thresholds due to a more compact representation of the data set. TreeMiner exceeds the 256GB system memory and can only fit into memory only 0.6\% of the Twitch data set. This is again mainly due to the high number of events $|\eventspace|$ which negatively affects the memory efficiency of TreeMiner. 

In the Spotify data set, the trie models show more than two orders of magnitude memory saving over the vector model of PrefixSpan. In particular, PrefixSpan uses approximately 10GB of memory while TreeMiner uses 0.12GB of memory with an average improvement of 98.8\%. The $\btrie$Miner and $\htrie$Miner algorithms use less than 0.02GB of memory and show improvements of 99.7\% over PrefixSpan and 78\% over TreeMiner. These results are mainly due to the low number of events $|\eventspace|$ in the Spotify data set, which allows higher compression in trie models. The $\btrie$Miner and $\htrie$Miner are equivalent in the Spotify data set as the hybrid algorithm determined the transition length as $\optlen = \maxseqsize$, giving a full trie model of the data set for $\htrie$Miner. Both algorithms are more efficient than TreeMiner due to their lower memory overhead at each trie node. 

The $\btrie$Miner and $\htrie$Miner algorithms enjoy the most significant improvements in large data sets. In the Criteo data set, PrefixSpan exceeds the 256GB of system memory and is only able to model 15\% of the data set. Although TreeMiner can model the entire Criteo data set, it exceeds system memory during its mining algorithm and terminates. This is in contrast to $\btrie$Miner and $\htrie$Miner, which use at least 2GB and at most 25GB of memory to mine the entire data set. This is more than an order of magnitude memory saving, and potentially amounts to 1.7TB lower memory usage than PrefixSpan requires to model the full data set into memory. This showcases the strength of trie models in modeling overlapping sequences using only a single path, while vector models use multiple vectors. Similar to the Spotify data set, $\btrie$Miner and $\htrie$Miner are equivalent with $\optlen = \maxseqsize$. 

The Genome data set is the most memory intensive, with $\htrie$Miner the only capable SPM algorithm within system memory. PrefixSpan exceeds system memory at 62\% of the Genome data set, TreeMiner at 3\%, and $\btrie$Miner at 29\%. This indicates a lower sequence overlap in the Genome data set, which gives an advantage to $\htrie$Miner over $\btrie$Miner.

Overall, $\htrie$Miner is the most memory-efficient SPM algorithm, closely followed by $\btrie$Miner. TreeMiner is the least memory-efficient algorithm, using considerably higher memory than all other algorithms and often failing to model medium to large data sets within system memory. While less efficient, the memory usage of PrefixSpan is still within the range of what is typical for regular PCs for small to medium data sets. However, this is not the case for large data sets, with $\htrie$Miner being the only capable SPM algorithm. We conclude that $\htrie$Miner provides compelling benefits in memory efficiency for any data set size over all other alternatives.

\subsection{Computational Time}\label{sec:results:time}

\begin{figure}
     \centering
         \captionsetup{justification=centering}
     \begin{subfigure}[b]{\textwidth}
         \centering
         \includegraphics[width=0.75\textwidth]{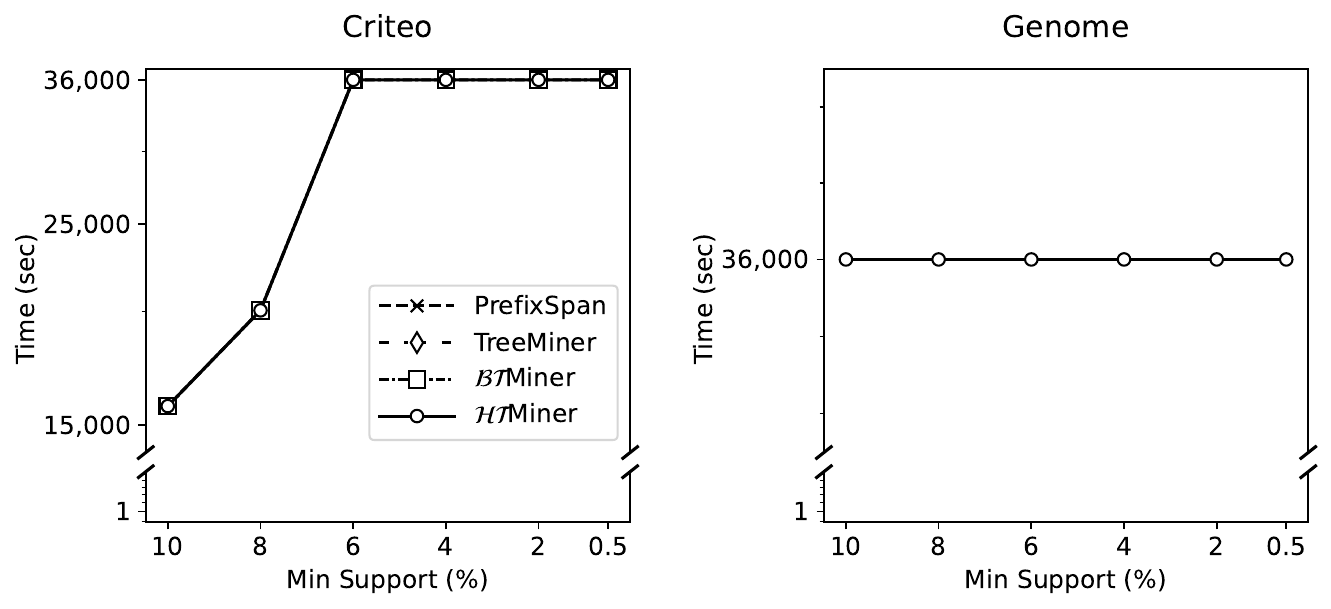}
         \caption{Large data sets. $\btrie$Miner and $\htrie$Miner exceed the imposed time limit of 36000 seconds when mining below the 6\% minimum support threshold for the Criteo data set. $\htrie$Miner exceeds the imposed time limit when mining the Genome data set at all thresholds. PrefixSpan and TreeMiner cannot model and mine either data set within system memory and are not reported for computational time.}
     \end{subfigure}
     \quad
     \begin{subfigure}[b]{\textwidth}
         \centering
         \includegraphics[width=0.75\textwidth]{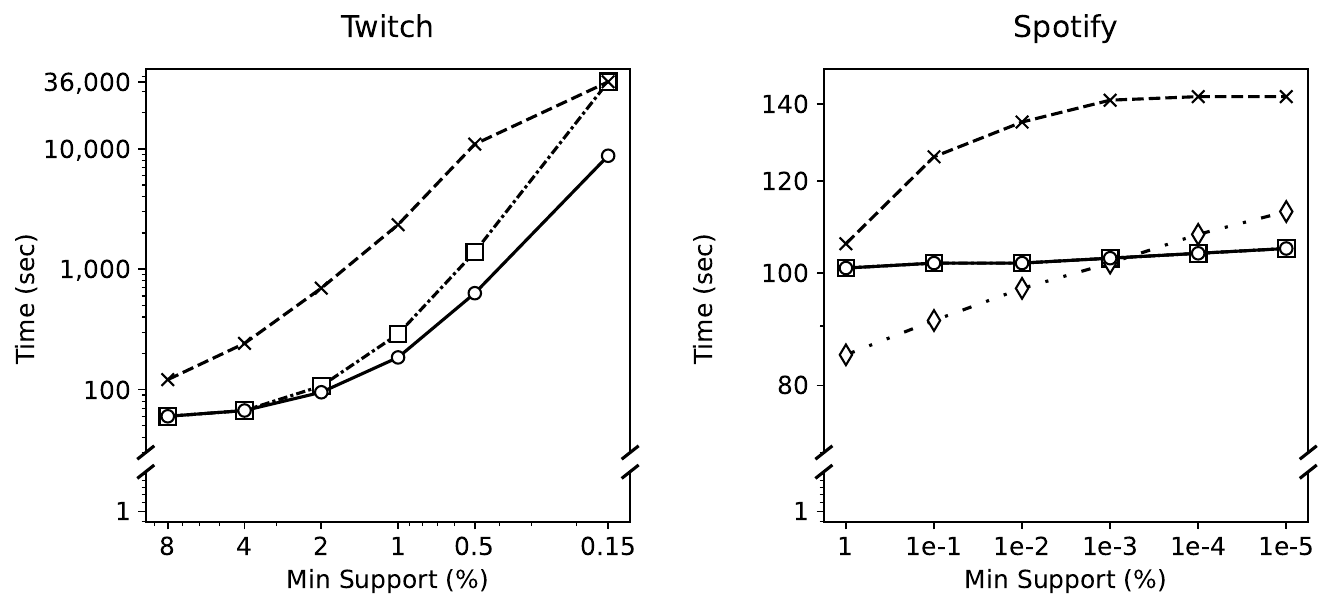}
         \caption{Medium data sets. TreeMiner cannot mine the Twitch data set within system memory and is not reported for computational time.}
     \end{subfigure}
          \quad
     \begin{subfigure}[b]{\textwidth}
         \centering
         \includegraphics[width=0.75\textwidth]{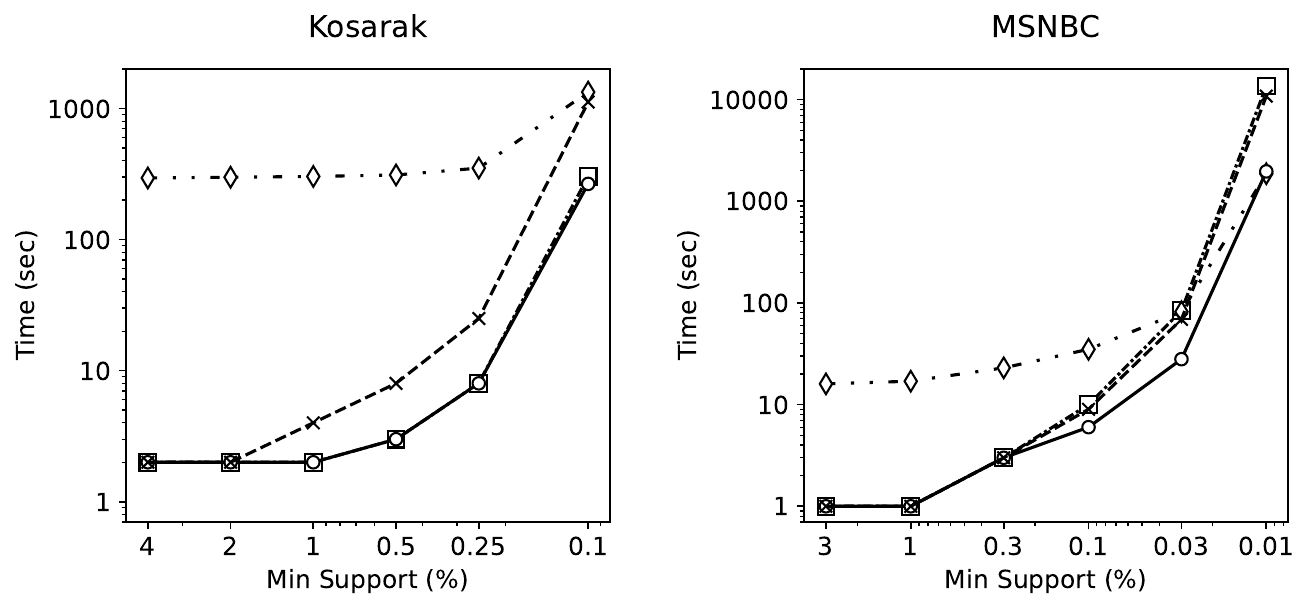}
         \caption{Small data sets.}
     \end{subfigure}
        \caption{Computation time.}
        \label{fig:comptime}
\end{figure}

Figure~\ref{fig:comptime} gives the computational time of all algorithms. The computational time of any algorithm that exceeds system memory is irrelevant and not reported. 

In the small data set Kosarak, TreeMiner is the slowest algorithm, especially for higher support thresholds. This is mainly due to the time required to populate its trie labels, which turned out to be costly in both time and memory efficiency. All other algorithms are mostly the same for higher support thresholds---where fewer patterns are mined. However, the improvements of $\htrie$Miner and $\btrie$Miner over PrefixSpan become larger for smaller thresholds, with more than a 60\% speedup (approximately 800 seconds) for both algorithms. Both algorithms show an average improvement of 95\% over TreeMiner, emphasizing the benefits in computational time provided by a less memory intensive trie model.

For the small data set MSNBC, all algorithms are mostly similar in time efficiency on higher support thresholds, but significantly different on lower support thresholds. This indicates that all algorithms are similar in constructing their data set representation, but not in mining it. PrefixSpan and $\btrie$Miner require approximately 11,000 seconds to mine the MSNBC data set at the lowest support threshold (mining more than 100,000,000 patterns). In contrast, TreeMiner and $\htrie$Miner show close to an order of magnitude faster mining, requiring approximately 1,800 seconds. The faster mining of TreeMiner shows the possible benefits of a richer set of trie labels, which increases mining efficiency when it can be efficiently modeled and fitted into memory. The faster mining of $\htrie$Miner shows the benefit of a hybrid data structure in the mining process---equaling the benefits provided by richer trie labels in time efficiency while consuming less memory. 

In the medium-size Twitch data set, $\btrie$miner and $\htrie$Miner show an average speedup of 50\% over PrefixSpan. Both the PrefixSpan and $\btrie$Miner reach the imposed time limit for the lowest support threshold, where PrefixSpan finds 15\% of total patterns and $\btrie$Miner finds 65\% of total patterns. $\htrie$Miner can mine the entire data set in approximately 9,000 seconds, showing at least a 75\% speedup. These show the benefits of mining trie paths that model multiple sequences over mining vector representations which model each sequence separately. In particular, the mining algorithm can mine multiple sequences in a pass over a single trie path $(\rootnode, \dots, \node)$, which requires $\freq_{\node}$ more passes in vector models to mine the same sequences. When $\freq_{\node}$ is lower, such as in the postfixes of the longer sequences of the MSNBC data set, $\htrie$Miner becomes more time-efficient than $\btrie$Miner.

Results on the Spotify data set show similar computational time on all algorithms, with at most a $[20-40]$ seconds difference. TreeMiner is slightly faster (20 seconds) in higher support thresholds, while $\htrie$Miner and $\btrie$Miner are slightly faster (10 seconds) for lower support thresholds. PrefixSpan is consistently slower (5-40 seconds), with the difference highest for small support thresholds. This is mainly to the high compression of trie models given the low number of events $|\eventspace|$ in this data set. Interestingly, $\htrie$Miner and $\btrie$Miner stay relatively constant in computational time over all support thresholds. This shows that most patterns are shorter in length, and demonstrates the efficiency of using ancestor labels in fast pruning of pattern extensions. 

The $\btrie$Miner and $\htrie$Miner algorithms show high computational time in large data sets. In the Criteo data set, both algorithms reach the imposed 36,000 seconds time limits for support thresholds lower or equal to 6\%. The computational time was considerably high for the Genome data set, where $\htrie$Miner reached the time limit at all thresholds. In particular, we observed slower than usual mining speed in the Genome data set as memory consumption was near system limits, increasing the overhead of memory access by the CPU. This shows that it is time-efficient to reduce memory consumption even when operating within---but close to---system limits. 

Overall, we observe that $\htrie$Miner is also faster than the state of the art, especially for larger data sets. Note that this is generally not the case for memory-efficient algorithms, as memory and time efficiency are often a trade-off. For both time and memory efficiency, we thus conclude that $\htrie$Miner is superior to all other SPM alternatives, providing considerably higher memory efficiency and often coupled with higher time efficiency.

\section{Conclusion}\label{sec:conclusion}

This paper develops a memory-efficient SPM algorithm using trie models of the data set. Our methodology involves a new binary trie model $\btrie$ that stores minimal information at its nodes to compactly represent the data set in memory. We show that this compact representation is sufficient for the subsequent SPM task, and develop a novel $\btrie$Miner algorithm to mine all sequential patterns. We build on our trie model to develop a hybrid model $\htrie$ that models prefixes with high overlap using a trie model, and postfixes with low overlap using a vector model. We integrate $\btrie$Miner and prefix-projection to develop $\htrie$Miner, which can effectively mine $\htrie$. We proved that $\htrie$Miner is always smaller and more memory efficient than pure vector or trie models and their corresponding algorithms. 

Numerical results on real-life test instances showed that on small and medium data sets, $\htrie$Miner provides, on average, 79\% (approximately 5GB) and 90\% (approximately 19GB) memory savings compared to the state-of-the-art vector and trie models, respectively. Furthermore, $\htrie$Miner was shown to be the only SPM algorithm capable of mining large data sets within 256GB of system memory, potentially saving 1.7TB in memory consumption. While memory efficiency is often a trade-off with time efficiency, $\htrie$Miner also showed lower computational time than the state of the art, with an average improvement of 43\% over vector models and 54\% over trie models.  

\bibliographystyle{plainnat2}
\bibliography{EFFSPM}

\end{document}